\documentclass[]{emulateapj}

\usepackage[colorlinks,urlcolor=blue,citecolor=blue,linkcolor=blue]{hyperref}
\slugcomment{DRAFT \today}
\shorttitle{New RCBs and DYPers from ACVS}
\shortauthors{Miller et.\ al.}

\begin{document}

\title{Discovery of Bright Galactic R Coronae Borealis and DY Persei Variables: Rare Gems Mined from ACVS}

\author{A.~A.~Miller\altaffilmark{1,*}, 
J.~W.~Richards\altaffilmark{1,2},
J.~S.~Bloom\altaffilmark{1},
S.~B.~Cenko\altaffilmark{1},
J.~M.~Silverman\altaffilmark{1},
D.~L.~Starr\altaffilmark{1},
and K.~G.~Stassun\altaffilmark{3,4}
}

\altaffiltext{1}{Department of Astronomy, University of California, Berkeley, CA 94720-3411, USA}
\altaffiltext{2}{Statistics Department, University of California, Berkeley, CA, 94720-7450, USA}
\altaffiltext{3}{Department of Physics and Astronomy, Vanderbilt University, Nashville, TN 37235, USA}
\altaffiltext{4}{Department of Physics, Fisk University, 1000 17th Ave. N., Nashville, TN 37208, USA}
\altaffiltext{*}{E-mail: {\tt amiller@astro.berkeley.edu}}

\begin{abstract}

We present the results of a machine-learning (ML) based search for new R Coronae Borealis (RCB) stars and DY Persei-like 
stars (DYPers) in the Galaxy using cataloged light curves from the All-Sky Automated Survey (ASAS) Catalog of Variable 
Stars (ACVS). RCB stars---a 
rare class of hydrogen-deficient carbon-rich supergiants---are of great interest owing to the insights they can provide on 
the late stages of stellar evolution. DYPers are possibly the low-temperature, low-luminosity analogs to the RCB phenomenon, 
though additional examples are needed to fully establish this connection. While RCB stars and DYPers are traditionally 
identified by epochs of extreme dimming that occur without regularity, the ML search framework more fully captures the 
richness and diversity of their photometric behavior. We demonstrate that our ML method can use newly discovered RCB stars 
to identify additional candidates within the same data set. Our search yields 15 
candidates that we consider likely RCB stars/DYPers: new spectroscopic observations confirm that four of these 
candidates are RCB stars and four are DYPers. Our discovery of four new DYPers increases the number of known 
Galactic DYPers from two to six; noteworthy is that one of the new DYPers has a measured parallax and is $m \approx 7$ mag, 
making it the brightest known 
DYPer to date. Future observations of these new DYPers should prove instrumental in establishing the RCB connection. We 
consider these results, derived from a machine-learned probabilistic classification catalog, as an important 
proof-of-concept for the efficient discovery of rare sources with time-domain surveys.

\end{abstract}


\section{Introduction}

R Coronae Borealis (RCB) stars are hydrogen-deficient carbon (HdC) stars that exhibit spectacular 
($\Delta{m}_V$ up to $\sim$8 mag), aperiodic declines in brightness (for a review on RCB stars see \citealt{Clayton96}). 
The fading occurs rapidly ($\sim$1 to few weeks) as new dust is formed in the circumstellar environment, 
and the recovery is slow, sometimes taking several years, as the new dust is dispersed and removed from the line of sight.
At maximum light RCB stars are bright supergiants, which in combination with the large-amplitude photometric 
variability should make them easy to discover. Yet, to date there are only $\sim$56 known RCB stars in the Galaxy 
\citep{Clayton96,Clayton02,Zaniewski05,Tisserand08,Clayton09,Kijbunchoo11}. The rarity of these stars suggests that 
they reflect a very brief phase of stellar evolution, or a bias in RCB star search methods, or both. 

The lack of hydrogen and overabundance of carbon in RCB atmospheres implies that RCB stars are in a late stage of stellar 
evolution, but no consensus has yet emerged regarding their true physical nature. There are two leading theories for 
explaining the observed properties of RCB stars: the Double Degenerate (DD) 
scenario and the final Helium shell flash (FF) scenario (see e.g., \citealt{Iben96}). The DD 
scenario posits that RCB stars are the stellar remnant of a carbon--oxygen white dwarf (WD) and 
helium WD merger. In the FF scenario, a thin layer of He in the interior of the star begins runaway burning, which leads to 
the rapid expansion of the photosphere shortly before the star becomes a planetary nebula. There are observational 
properties of RCB stars that both theories have difficulty explaining \citep{Clayton96}, and conflicting observational 
evidence supporting aspects of both (e.g., \citealt{Clayton07,Pandey08,Clayton06,Clayton11}).
If, as some of the recent observations suggest, the DD scenario proves correct, then a 
complete census of Galactic RCB stars should be able to calibrate population synthesis models of WD binary systems 
(e.g., \citealt{Nelemans01}), which may improve our understanding of these systems as the progenitors of Type Ia supernovae. 
In any event, the enigmatic nature of these rare objects, and the opportunity to elucidate the astrophysics of an important 
late stage of stellar evolution, motivates us to search for additional benchmark exemplars of the class.

Based on the detection of RCB stars in the Large Magellanic Cloud (LMC), it is argued in \citet{Alcock01} that there 
should be $\sim$3200 RCB stars in the Galaxy. With the actual number of known RCB stars in the Milky Way roughly two 
orders of magnitude below this estimate, this suggests that either thousands of RCB stars 
remain undetected or the differing star formation 
environments/histories in the LMC and the Milky Way result in highly different RCB populations. An observational bias that 
preferentially selects warm RCB stars likely contributes to the discrepancy between the predicted and known number of these 
stars in the Galaxy \citep{Lawson90b}. Indeed, recent discoveries of RCB stars in the Galactic bulge and Magellanic Clouds 
(MCs) have uncovered more cool, $T_{\rm eff} \sim$5000 K, rather than warm, $T_{\rm eff} \sim$7000 K, RCB stars 
\citep{Alcock01,Zaniewski05,Tisserand08,Tisserand09}. The observed correlation between color 
and $M_V$, with bluer RCB stars in the MCs being more luminous \citep{Alcock01,Tisserand09}, 
clearly shows that any magnitude-limited survey will have an observational bias towards discovering the 
intrinsically rarer warm RCB stars. 

There may also be a large population of RCB stars that have colder photospheres than the cool RCB stars: there is one 
known Galactic RCB star, DY Persei \citep{Alksnis94}, that has $T_{\rm eff} \sim$3500 K \citep{Keenan97}. Recent 
observations of the MCs have identified several DY Persei-like stars (DYPers) while searching for RCB stars 
\citep{Alcock01,Tisserand09,Soszynski09}, while \citet{Tisserand08} 
discovered the second known DYPer in the Milky Way using observations of the Galactic bulge. In addition to cooler 
photospheres, DYPers have other properties that differ from RCB stars, which has led to some degree of ambiguity regarding 
the connection between these two classes (see e.g., \citealt{Alcock01,Tisserand09,Soszynski09}). 

DYPers and RCB stars both show an overabundance of carbon in their atmospheres and unpredictable, large-amplitude declines in 
their light curves. Several properties differ between the two, however, for instance, DYPers: (i) have symmetric declines 
in their light curves, (ii) clearly show $^{13}$C in their spectra, (iii) are on average $\sim$10 times fainter than 
RCB stars, and (iv) may have significant H in their atmospheres. A detailed examination of the differences in the 
mid-infrared excesses of RCB stars and DYPers in the MCs led to the conclusion in \citet{Tisserand09} that DYPers are 
most likely normal carbon stars that experience ejection events rather than an extension of the RCB phenomenon 
to lower temperature stars. Furthermore, using OGLE-III observations, it is shown in \citet{Soszynski09} that 
several carbon-rich asymptotic giant 
branch stars (AGBs), which have been classified as Mira or semi-regular periodic variables on the basis of their light 
curves, show evidence for DYPer-like declines in their light curves.\footnote{We note that the sources included in the 
study of \citet{Soszynski09} are photometrically classified as carbon AGB. Thus, the candidates in that  
study require spectroscopic observations in order to be confirmed as DYPers.} This leads to the conclusion in 
\citeauthor{Soszynski09} that DYPers are heavily enshrouded carbon-rich AGB stars that are an extension of typical 
variables rather than a separate class of variable stars. Nevertheless, all 
studies of DYPers to date have cited a need for more observations, in particular high resolution spectra to conduct 
detailed abundance analyses, to confirm or deny the possibility that DYPers are the low temperature analogs to RCB stars.

Over the past decade the decrease in the cost of large CCDs, coupled with a dramatic increase in computer processing 
power and storage capabilities, has enabled several wide-field, time-domain surveys. These surveys will continue 
to produce larger data sets before culminating near the end of the decade with the Large 
Synoptic Survey Telescope (LSST; \citealt{Ivezic08}). This explosion of observations should enable the discovery of the 
thousands of ``missing'' Galactic RCB stars, should they in fact exist. These new discoveries do not come without a 
cost, however, as the data rates of astronomical surveys are now becoming enormous. While it was once feasible for humans 
to visually examine the light curves of all the newly discovered variable stars, as the total number of photometric variables 
grows to 10$^6$--10$^7$ visual inspection by expert astronomers becomes intractable. 

Advanced software solutions, such as machine-learning (ML) 
algorithms, are required to analyze the vast amounts of data produced by current and upcoming time-domain surveys.  In an ML 
approach to classification, data from sources of known science class are employed to train statistical algorithms 
to automatically learn the distinguishing characteristics of each class. These algorithms generate an 
optimal predictive model that can determine the class (or posterior class probability) of a new source given its observed 
data.\footnote{For a primer on machine learning, we refer the reader to \citet{Hastie09}.} \citet{richards11} 
presented an end-to-end ML framework for multi-class variable star classification, in which they describe algorithms for 
feature generation from single-band light curves and outline a methodology for non-parametric, multi-class statistical 
classification.  

In this paper we present the results of a search for new RCB stars and DYPers in the Galaxy using version 2.3 of the ML 
catalog presented in \citet{Richards12b}. In \S 2 we describe 
the candidate selection procedure, while \S 3 describes the new and archival observations of the candidates. Our analysis of 
the photometric and spectroscopic data is contained in \S 4. The individual stars are examined in further detail in \S 5, 
while we discuss the results in \S 6. Our conclusions are presented in \S 7. 

\section{Candidate Selection}\label{sec:cands}

\subsection{Advantages of Machine-Learning Classification}\label{sec:ML}

Candidate selection of possible RCB stars was performed using version 2.3 of the machine-learned ACVS classification catalog 
(\href{http://www.bigmacc.info/}{MACC}; \citealt{Richards12b}) of variable sources cataloged from All-Sky Automated Survey (ASAS; \citealt{Pojmanski97, 
Pojmanski01}). Full details of the classification procedure can be found in \citet{Richards12} and \citet{Richards12b}. 
Briefly, we employ a Random Forest (RF) classifier, which has been shown to provide the most robust results for 
variable star classification (see e.g., \citealt{richards11,Dubath11}), to provide probabilistic classifications for all of 
the 50,124 sources in ASAS Catalog of Variable Stars (ACVS; \citealt{Pojmanski00}). The classification procedure 
proceeds as follows: for each source in the ACVS 71 features are computed, 66 from the ASAS light curves (e.g., period, 
amplitude, skew, etc.; for the full list of features we refer the reader to \citealt{Richards12b} and references therein) 
and 5 color features from optical and near-infrared (NIR) catalogs. A training set, upon which the RF classifications will be 
based, is constructed using light curves from 28 separate science classes, most of which are defined using well studied stars 
with high precision light curves from the {\it Hipparcos} and OGLE surveys (\citealt{debosscher07,richards11}), as well as some 
visually classified sources from ACVS for a few of the classes that are not well represented in {\it Hipparcos} 
\citep{Richards12}. The same 71 features are calculated for all the sources in the training set, and the RF classifier uses 
the separation of the 28 science classes in the multi-dimensional feature space to assign probabilistic classifications to 
each source in the ACVS. In the end, the probability of belonging to each individual science class is provided for each ACVS 
source and a post-RF procedure is used to calibrate these probabilities (meaning that a source with $P({\rm Mira}) = 0.5$ 
has a $\sim$50\% chance of actually being a Mira). 

When searching for RCB stars in time-domain survey data, RF classification provides a number of advantages 
relative to the more commonly used method of placing hard cuts on a limited set of a few features. Many studies have 
focused on light curves with large amplitude variations and a lack of periodic signal (e.g., \citealt{Alcock01, 
Zaniewski05,Tisserand08}). A few recent studies have noted that additional cuts on NIR and mid-infrared colors can improve 
selection efficiency \citep{Soszynski09,Tisserand11,Tisserand12}. While these surveys have all proven successful, the 
use of hard cuts may eliminate actual RCB stars from their candidate lists.

Hard cuts are not necessary, however, when using a multi-feature RF classifier, which is capable (in principle) of capturing 
most of the photometric behavior of RCB stars (including the large-amplitude, aperiodic fades from maximum light as well as 
the periodic variations that occur near maximum light). Another 
general disadvantage in the use of hard cuts for candidate selection of rare sources is that the hard cuts are 
typically defined by known members of the class of objects for which the search is being conducted. Any biases 
present in the discovery of the known members of a particular class will then be encoded into the absolute (i.e., hard 
cuts) classification schema. This can exclude subclasses of sources that differ slightly from the defining members of 
a class. Furthermore, new discoveries will be unable to refine the selection criteria since, by construction, 
they will fall within the same portion of feature space as previously known examples. 

The RF classifier produces an estimate of the posterior probability that a source is an RCB star given its light curve and 
colors. This allows us to construct a relative ranking of the RCB likelihood for all the sources in ACVS.  
Instead of making cuts in feature space, we can search down the ordered list of candidates. In this sense the 
RF classifier identifies the sources that are closest to the RCB training set relative to the other classes. The RF 
classifier finds the class boundaries in a completely data-driven way, allowing for the optimal use of known 
objects to search for new candidates in multi-dimensional feature spaces. This helps to mitigate against biases present 
in the training set, as 
classifications are performed using the location of an individual source in the multidimensional feature phase-space 
volume relative to defined classes in the training set. 

\subsection{The Training Set}\label{sec:training_set}

The MACC RCB training set was constructed using high-confidence positional matches between ACVS sources and known RCB stars 
identified in SIMBAD\footnote{{\tt http://simbad.u-strasbg.fr/simbad/}} and the literature. In total there are 18 cataloged 
RCB stars that are included in the ACVS, which we summarize in Table~\ref{tab:training_set}. The light curves of the known RCB 
stars were visually examined for the defining characteristic of the class: sudden, aperiodic drops in brightness followed by 
a gradual recovery to pre-decline flux levels. All of the known RCB stars but one, ASAS 054503$-–$6424.4, showed evidence for 
such behavior. ASAS 054503$-–$6424.4 is one of the brightest RCB stars in the LMC ($V_{\rm max} \approx 13.75$ mag), which 
during quiescence is barely above the ASAS detection threshold. The light curve for ASAS 054503$-–$6424.4 does not show a 
convincing decline from maximum light, and as such we do not include it in the training set. 

In addition to the 18 RCB 
stars in ACVS, 7 additional RCB stars are detected in ASAS with the characteristic variability of the class.\footnote{They 
are: SU Tau, UX Ant, UW Cen, V348 Sgr, GU Sgr, RY Sgr, and V532 Oph.} These sources all have clearly variable ASAS light 
curves; their exclusion from the ACVS means there is some bias in the construction of that catalog. In order to keep this 
bias self-consistent the training set for the MACC only included sources from \citet{richards11} and supplements from 
ACVS (see \citealt{Richards12}). We note that a future paper to classify all $\sim$12 million sources 
detected by ASAS will include all ASAS RCB stars in its training set (Richards et al., in prep). Therefore the training set 
includes 17 RCB stars, which is limited by the coverage and depth of ASAS, the selection criteria of the ACVS, and the 
paucity of known RCB stars in the Galaxy. There are no known DYPers in 
ACVS: only two are known in the Galaxy and the DYPers in the MCs are fainter than the ASAS detection limits. 
Nevertheless, the similarity in the photometric behavior of RCB stars and DYPers allows us to use the RCB training set 
to search for both types of star. As more Galactic RCB stars and DYPers are discovered, we will be able to supplement 
the training set and improve the ability of future iterations of the RF classifier (see \S 6.2). 

\begin{deluxetable*}{lllccr}
\tablecolumns{6}
\tabletypesize{\tiny}
%\rotate
\tablewidth{0pc} 
\tablecaption{Known RCB Stars in ACVS.} 
\tablehead{ 
\colhead{Name} & \colhead{Other ID} & \colhead{DotAstro\tablenotemark{a}} & \colhead{Training\tablenotemark{b}} & \colhead{$P({\rm RCB})$\tablenotemark{c}} & \colhead{$R({\rm RCB})$\tablenotemark{d}} \\
\colhead{} &  \colhead{} & \colhead{ID} & \colhead{Set?} & \colhead{} & \colhead{}
} 
\startdata 
ASAS 054503$-$6424.4 & HV 12842 & 220040 & N & 0.010\tablenotemark{e} & 2806\tablenotemark{e} \\
ASAS 143450$-$3933.5 & V854 Cen & 240306 & Y & 0.914 & 2 \\
ASAS 150924$-$7203.8 & S Aps & 241463 & Y & 0.934 & 2 \\
ASAS 154834+2809.4 & R CrB & 242999 & Y & 0.430 & 34 \\
ASAS 162419$-$5920.6 & RT Nor & 244506 & Y & 0.629 & 10 \\
ASAS 163242$-$5315.6 & RZ Nor & 244888 & Y & 0.944 & 2 \\
ASAS 171520$-$2905.6 & V517 Oph & 247066 & Y & 0.169 & 158 \\
ASAS 172315$-$2252.0 & V2552 Oph & 247575 & Y & 0.105 & 425 \\
ASAS 180450$-$3243.2 & V1783 Sgr & 250762 & Y & 0.642 & 9 \\
ASAS 180850$-$3719.7 & WX CrA & 251121 & Y & 0.834 & 2 \\
ASAS 181325$-$2546.9 & V3795 Sgr & 251489 & Y & 0.726 & 8 \\
ASAS 181509$-$2942.5 & VZ Sgr & 251638 & Y & 0.845 & 2 \\
ASAS 181851$-$4632.9 & RS Tel & 251987 & Y & 0.951 & 1 \\
ASAS 184732$-$3809.6 & V CrA & 254404 & Y & 0.804 & 3 \\
ASAS 190812+1737.7 & SV Sge & 256072 & Y & 0.053 & 859 \\
ASAS 191012$-$2029.7 & V1157 Sgr & 256221 & Y & 0.276 & 74 \\
ASAS 193222$-$0011.5 & ES Aql & 257713 & Y & 0.044 & 901 \\
ASAS 220320$-$1637.6 & U Aqr & 263740 & Y & 0.839 & 3 \\
\enddata
\tablenotetext{a}{ID from the MACC.}
\tablenotetext{b}{Indicates whether or not (Y/N) this star was included in the RCB training set.}
\tablenotetext{c}{Calibrated probability of belonging to the RCB class obtained when source is left out of the training set for cross validation.}
\tablenotetext{d}{Relative rank of RCB likelihood when source is left out of the training set for cross validation.}
\tablenotetext{e}{Source is not included in the training set, values taken directly from the MACC.}

\label{tab:training_set}
\end{deluxetable*}%
 
In order to determine our ability to recover known RCB stars using the RF classifier we perform a leave-one-out cross 
validation (CV) procedure. For the 17 sources in the RCB training set, we remove one source and re-run the RF classifier 
in an identical fashion to that used in \citet{Richards12b}. We then record the RF-determined probability that 
the removed source belongs to the RCB class, $P(\rm RCB)$, and the ranked value of $P(\rm RCB)$ relative to all other stars 
that are not included in the training set, $R({\rm RCB})$. We repeat the CV procedure for each star included in the 
training set, and the results are shown 
in Table~\ref{tab:training_set}. Since the training set is being altered in each run of the CV, $R({\rm RCB})$ provides a 
better measure of the quality of each candidate; $R({\rm RCB})$ is a relative quantity, whereas the calibration of 
$P(\rm RCB)$ will differ slightly from run to run. Eight of the 17 sources in the training set have $R({\rm RCB}) \leq 3$, 
implying that $\sim$50\% of the training set would be a top three candidate RCB star had we previously not known about it. 
Fifteen of the 17 RCB stars in the training set would be in the top 0.8\% of the 50,124 sources in the ACVS, while all the 
known RCB stars in ACVS, including ASAS 054503$-$6424.4 (which is not in the training set), are in the top $\sim$6\% of RCB 
candidates. Two sources in the training set, SV Sge and ES Aql, are not listed near the top of $R({\rm RCB})$ ranking during 
CV. For ES Aql this occurs 
because the star is highly active during the ASAS observations showing evidence for at least six separate declines during the 
$\sim$10 yr observing period. As a result the light curve folds fairly well on a period of $\sim$397 day, and ES Aql becomes 
confused with Mira and semi-regular periodic variables (see Figure~\ref{fig:hardcuts}). SV Sge, on the other hand, shows 
significant periodicity at the parasite frequency of 1 day, which precludes it from having a high $R({\rm RCB})$.
The CV procedure allows us to roughly tune the efficiency of our selection criteria; the purity of the selection criteria 
cannot be evaluated until candidates have been spectroscopically confirmed.

\subsection{The Candidates}\label{sec:macc_cands}
Due to the relative rarity of RCB stars, we elected to generate a candidate list with high efficiency while 
sacrificing the possibility of high purity. With only $\sim$50 Galactic RCB stars known to date, every new discovery has the 
potential to add to our knowledge of their population and characteristics. To generate our candidate list, we selected all 
sources from the MACC with $P({\rm RCB}) >$ 0.1, 
which resulted in a total of 472 candidates. The selection criterion was motivated by the CV experiment, 
which indicates that our candidate list should have an efficiency $\gtrsim$80\%. To obtain an efficiency close to 1 would 
require visual examination of roughly 3000 sources. 

Since the expected purity of our sample is small by design, we examine the light curves of all sources within 
our candidate list by eye to remove sources that are clearly not RCB stars. These interlopers are typically  
semi-regular pulsating variables or Mira variables, often with minimum brightness levels below  
the detection threshold. We use the {\tt ALLSTARS} Web interface \citep{Richards12} to examine candidates, which 
in addition to light curves provides summary statistics (period, amplitude, color, etc.) for each source, as well 
as links to external resources, such as SIMBAD. We also remove any sources from the candidate list that are 
spectroscopically confirmed as non-carbon stars. 

\begin{figure}
\begin{center}
\includegraphics[width=80mm]{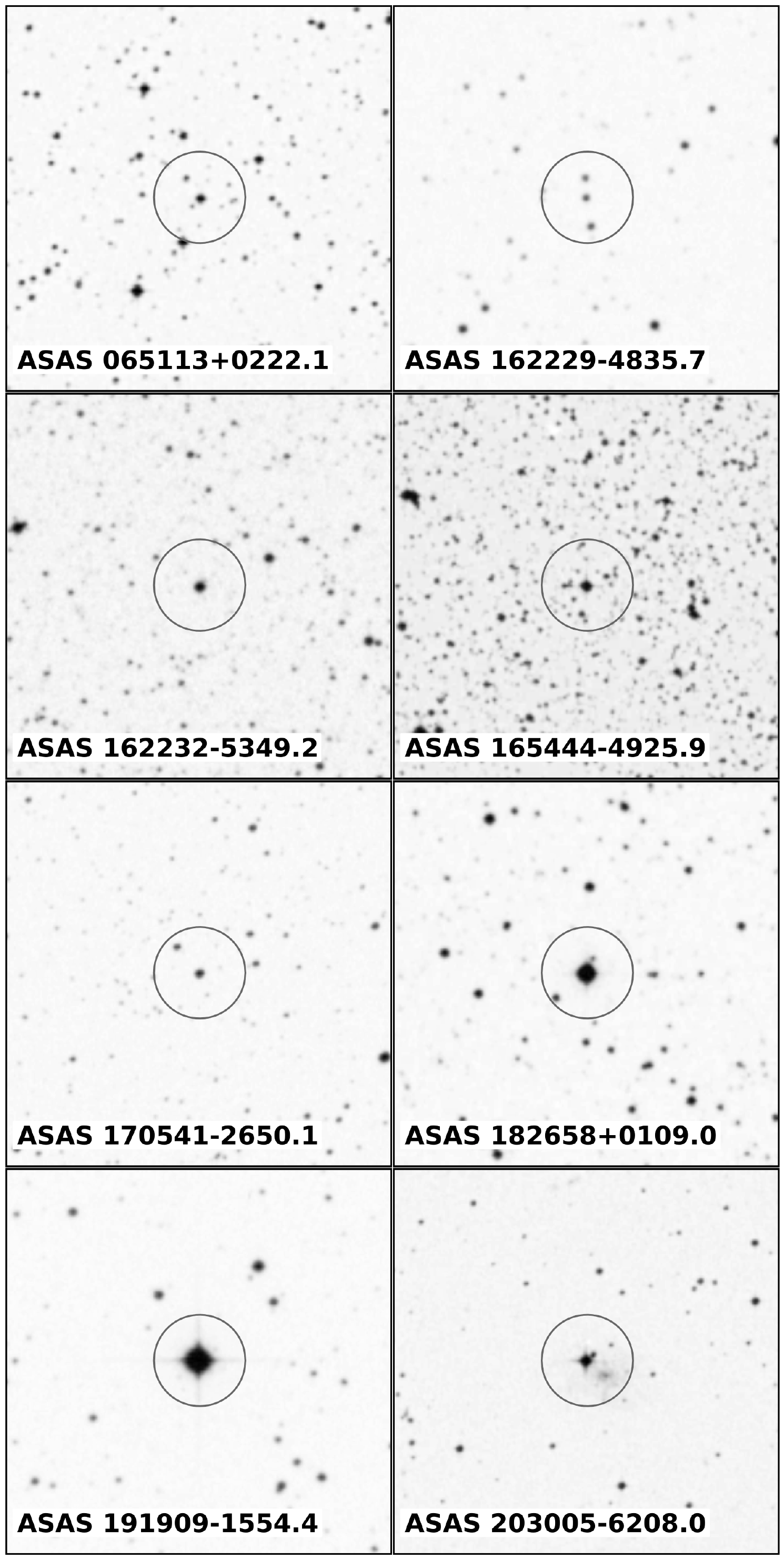}
\caption{Optical finding charts of the newly discovered RCB stars and DYPers. Each finder is $5\arcmin \times 
5\arcmin$ with north up and east to the left. The circles show the location of the targets and have $r = 33\farcs5$ which 
is the typical FWHM for ASAS images \citep{Pojmanski97}. The large pixels on the ASAS camera 
result in PSFs that include the light from several stars, meaning that some ASAS light curves underestimate the 
true variability of the brightest star within the PSF.}
\label{fig:RCBfinders}
\end{center}
\end{figure}

Following the removal of these stars the candidate list was culled from 472 to 15 candidates we considered likely RCB 
stars, for which we obtained spectroscopic follow-up observations. The general properties of the 15 spectroscopically 
observed candidates, including their names, coordinates, and RF probabilities, are summarized in Table~\ref{tab:coords}. 
Finding charts using images from the Digitized Sky Survey\footnote{\tt http://stdatu.stsci.edu/dss/} (DSS) for the 
spectroscopically confirmed RCB and DYPer candidates can be found in Figure~\ref{fig:RCBfinders}. Six of the 
selected candidates for spectroscopic observations are known carbon stars listed in the General Catalog of Galactic 
Carbon Stars (CGCS; \citealt{Alksnis01}; see Table~\ref{tab:coords}).

\begin{deluxetable*}{lllrrccrr}
\tablecolumns{9}
\tabletypesize{\tiny}
%\rotate
\tablewidth{0pc} 
\tablecaption{RCB Candidates with $P({\rm RCB}) >$ 0.1 from the MACC.} 
\tablehead{ 
\colhead{Name} & \colhead{Other ID} & \colhead{MACC\tablenotemark{a}} & \colhead{$\alpha_{\rm J2000.0}$\tablenotemark{b}} & \colhead{$\delta_{\rm J2000.0}$\tablenotemark{b}} & \colhead{CGCS\tablenotemark{c}} &   \colhead{$P({\rm RCB})$} & \colhead{$R({\rm RCB})$\tablenotemark{d}} & \colhead{R/D/N\tablenotemark{e}} \\
\colhead{} &  \colhead{} & \colhead{ID} & \colhead{(hh mm ss.ss)} & \colhead{(dd mm ss.s)} & \colhead{ID} & \colhead{} & \colhead{} & \colhead{}
} 
\startdata 
ASAS 060105+1654.7 & V339 Ori & 220556 & 06 01 04.65 & +16 54 40.8 &  ... & 0.466 & 25 & N \\
ASAS 065113+0222.1 & C* 596 & 223100 & 06 51 13.31 & +02 22 08.6 &  1429 & 0.302 & 73 & D \\
ASAS 073456$-$2250.1 & V455 Pup & 225801 & 07 34 56.24 & $-$22 50 04.2 &  1782 & 0.123 & 283 & N \\
ASAS 095221$-$4329.7 & IRAS 09503$-$4315 & 232170 & 09 52 21.37 & $-$43 29 40.5 &  ... & 0.617 & 10 & N \\
ASAS 153214$-$2854.4 & BX Lib & 242289 & 15 32 13.48 & $-$28 54 21.6 &  ... & 0.367 & 43 & N \\
ASAS 162229$-$4835.7 & IO Nor & 244409 & 16 22 28.84 & $-$48 35 55.8 &  ... & 0.950 & 1 & R \\
ASAS 162232$-$5349.2 & C* 2322 & 244411 & 16 22 32.08 & $-$53 49 15.6 &  3685 & 0.391 & 36 & D \\
ASAS 165444$-$4925.9 & C* 2377 & 245841 & 16 54 43.60 & $-$49 25 55.0 &  3744 & 0.490 & 22 & R \\
ASAS 170541$-$2650.1 & GV Oph & 246478 & 17 05 41.25 & $-$26 50 03.4 &  ... & 0.702 & 8 & R \\
ASAS 180823$-$4439.8 & V496 CrA & 251092 & 18 08 23.05 & $-$44 39 46.7  & ... & 0.110 & 389 & N \\
ASAS 182658+0109.0 & C* 2586 & 252675 & 18 26 57.64 & +01 09 03.1 &  4013 & 0.115 & 343 & D\\
ASAS 185817$-$3543.8 & IRAS 18549$-$3547 & 255280 & 18 58 17.19 & $-$35 43 44.7 &  ... & 0.127 & 251 & N \\
ASAS 191909$-$1554.4 & V1942 Sgr & 256869 & 19 19 09.60 & $-$15 54 30.1 &  4229 & 0.543 & 17 & D \\
ASAS 194245$-$2137.0 & ... & 258411 & 19 42 45.05 & $-$21 36 59.8 &  ... & 0.112 & 376 & N \\
ASAS 203005$-$6208.0 & NSV 13098 & 261023 & 20 30 04.96 & $-$62 07 59.2 &  ... & 0.340 & 52 & R \\
\enddata
\tablecomments{This table contains only those sources which were selected for spectroscopic follow-up following visual inspection of their light curves.}
\tablenotetext{a}{DotAstro ID: internal designation for the MACC.}
\tablenotetext{b}{Reported coordinates from the Two Micron All Sky Survey point source catalog \citep{cutri03}.}
\tablenotetext{c}{ID from the General Catalog of Galactic Carbon Stars (CGCS; \citealt{Alksnis01}).}
\tablenotetext{d}{Relative rank of $P({\rm RCB})$ including all sources from version 2.3 of the MACC not in the RCB training set.}
\tablenotetext{e}{Flag indicating classification of the source following spectroscopic observations: {\it R}: RCB, {\it D}: DYPer, {\it N}: Neither.}
\label{tab:coords}
\end{deluxetable*}%

\subsection{Feature Importance}

RF classifiers can provide quantitative feedback about the relative importance of each feature used for classification. The 
RF feature importance measure describes the decrease in the overall classifier performance following the replacement of a 
single feature with a random permutation of its values (see \citealt{Breiman01} for further details). We measure the 
importance of each feature using the average importance from a one-versus-one classifier whereby the RCB class is 
iteratively classified against each of the 27 other science classes on an individual basis. This procedure is run five times 
and the average of all runs is taken to reduce the variance present in any single run. Unsurprisingly, we find that  
\verb amplitude  is the most important feature. The importance measure does not properly capture the covariance between 
features and as a result the majority of the important features have to do with amplitude. The second and third most 
important features that are not highly covariant with amplitude are \verb1qso_log_chi2nuNULL_chi2nu1,\footnote{This 
is the same as the $\chi^2_{{\rm QSO},False}$ statistic, which is defined in \citet{Butler11}.} 
a measure of the dissimilarity between the photometric variations of the source and a typical quasar, and 
\verb freq1_harmonics_freq_0,  the best fit period. Interestingly, \verb freq_signif,  the 
significance of the best fit period of the light curve, ranks as only the 31st most important out of the 71 features. 

\begin{figure*}
\begin{center}
\includegraphics[width=180mm]{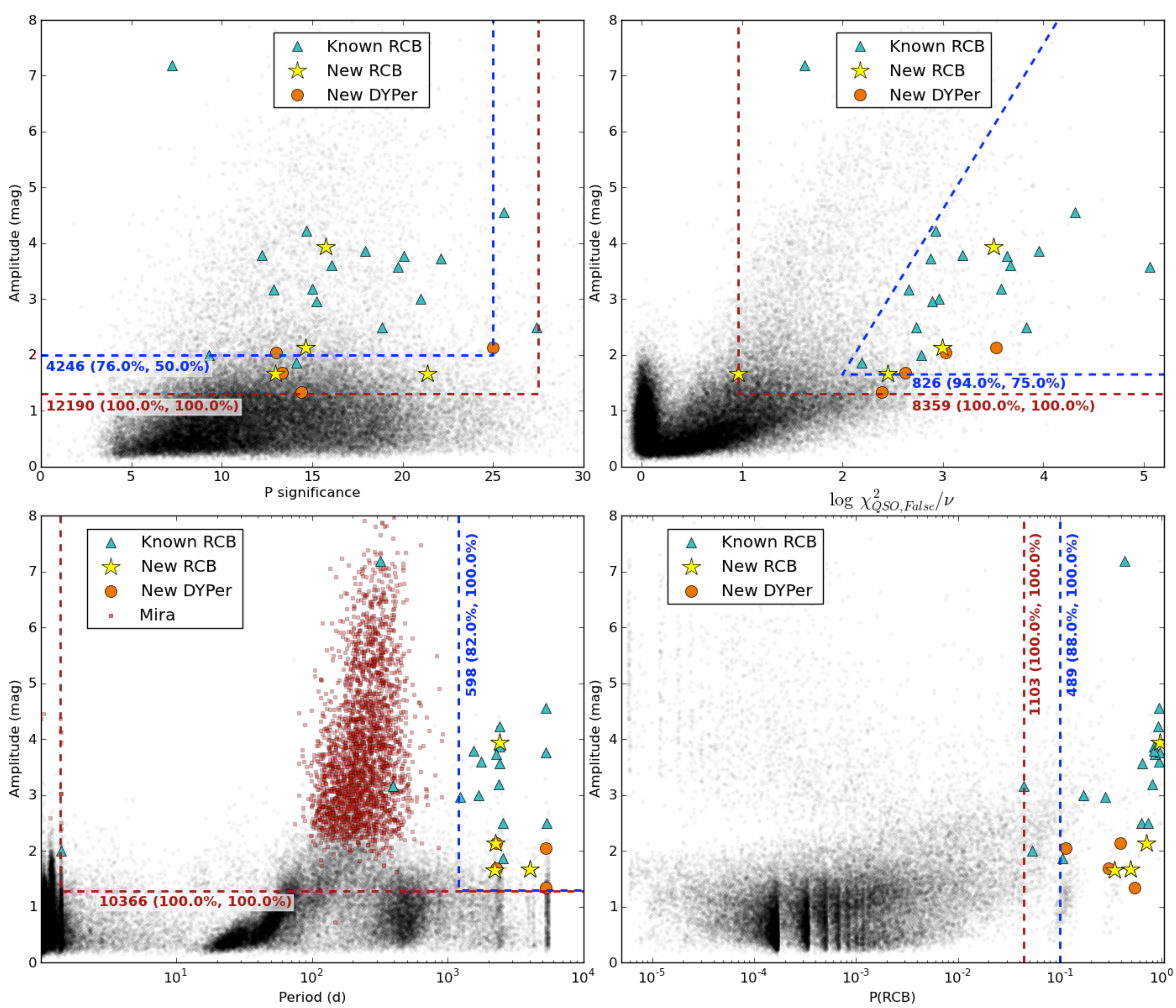}
\caption{Two dimensional cuts through the multi-dimensional feature space used to classify sources in version 2.3 of the MACC. 
Each panel 
shows the location of all sources in the MACC (black points), as well as the RCB stars in the training set (cyan triangles), 
newly discovered RCB stars (yellow stars), and new DYPers (orange circles). Also shown are cuts necessary to achieve 
$\sim$80\% (blue dashed line) and $\sim$100\% (red dashed line) RCB selection efficiency. Next to these lines are the total 
number of ACVS sources within the cut region, as well as the efficiency of recovering training set and new detections (shown 
in parenthesis), respectively. {\it Upper left}: amplitude versus period significance. {\it Upper right}: amplitude versus 
$\chi^2_{{\rm QSO},False}$. {\it Lower left}: amplitude versus period. Also shown is the tight cluster of Mira variables (red 
points), defined here as all ACVS sources with $P({\rm Mira}) > 0.7$. {\it Lower right}: amplitude versus $P({\rm RCB})$. 
Note that these are highly covariant as $P({\rm RCB})$ is strongly dependent on amplitude, which is why the cuts presented 
are shown in a single dimension. The $P({\rm RCB})$ values for the stars used in the training set are taken from the CV 
experiment from \S~\ref{sec:training_set}. }
\label{fig:hardcuts}
\end{center}
\end{figure*}

We summarize the results of these findings with two-dimensional cuts through the multi-dimensional feature space showing 
amplitude versus period significance, $\chi^2_{{\rm QSO},False}$, and period in Figure~\ref{fig:hardcuts}. We also show 
amplitude versus $P({\rm RCB})$. In each panel we show the location of the RCB stars in the training set as well as the 
newly discovered RCBs and DYPers presented in this paper, and we use the $P({\rm RCB})$ values from the CV experiment 
from \S 2.2 for the RCB stars in the training set. We also show the location of cuts necessary to achieve $\sim$80\% 
efficiency (blue dashed line) when selecting candidates using only two features, as well as the cuts necessary to achieve 
$\sim$100\% efficiency (red dashed line). As would be expected based on the results presented above, it is 
clear that $\chi^2_{{\rm QSO},False}$ and period are far more discriminating than period significance when selecting RCB 
candidates. To achieve an efficiency near 100\%, $P({\rm RCB})$ is vastly superior to any two dimensional slice through 
feature space. We note that the discretization seen in the distribution of $P({\rm RCB})$ is the result of using a 
finite number of trials 
within the RF classifier. The probability of belonging to a class is defined as the total number of times a source is 
classified within that class divided by the total number of trials. These discrete values are then smeared following the 
calibration procedure described in \S 2.1.

Many of the known and new RCB stars have very similar measured best periods clustered near $\sim$2400 and 5300 days, 
which for each corresponds to the largest period searched during the Lomb-Scargle analysis in \citet{Richards12b}. Folding 
these light curves on the adopted periods clearly shows that they are not periodic on the adopted periods, despite the relatively 
high period significance scores (see the upper left panel of Figure~\ref{fig:hardcuts}), which suggests some peculiarity in 
the feature generation process for these sources. We are exploring improved metrics for periodicity to be used in future 
catalogs. Nevertheless, despite these spurious period measurements, the ML classifier has correctly identified that this 
feature tends to be erroneous for RCB stars, and as such it is a powerful discriminant for finding new examples of the class. 

\section{Archival Data and New Observations}

\subsection{ASAS Photometry}

\begin{figure*}
\begin{center}
\includegraphics[width=130mm]{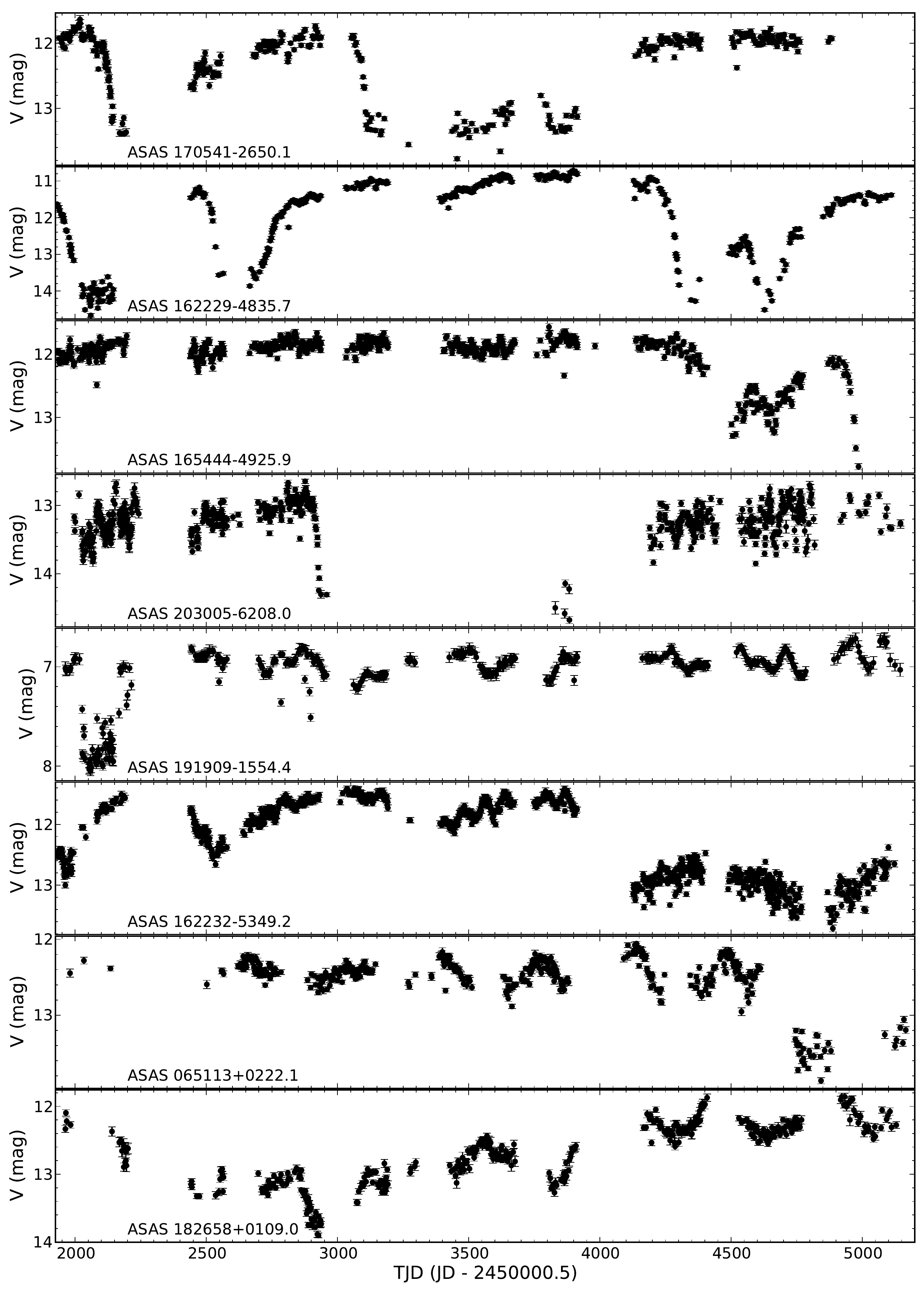}
\caption{ASAS $V$-band light curves of newly discovered RCB stars and DYPers. Note the differing magnitude ranges 
shown for each light curve. Spectroscopic observations confirm the top four candidates to be RCB stars, while the bottom 
four are DYPers.}
\label{fig:RCBlc}
\end{center}
\end{figure*}

All optical photometric observations were obtained during ASAS-3, which was an extension 
of ASAS, conducted at the Las Campanas Observatory (for further details on ASAS and ASAS-3 see 
\citealt{Pojmanski97,Pojmanski01}). Light curves were downloaded from the 
ACVS\footnote{\url{http://www.astrouw.edu.pl/asas/?page=acvs}}, and imported into our DotAstro.org 
(\url{http://dotastro.org}) astronomical light-curve warehouse for visualization and used with internal frameworks 
\citep{Brewer09}. The ACVS provides $V$-band measurements for a set of 50,124 pre-selected ASAS variables, 
measured in five different apertures of varying size \citep{Pojmanski02}. 
For each star in the catalog an optimal aperture selection procedure is used to determine the final light 
curve, as described in \citet{Richards12b}.\footnote{These optimal aperture light curves can be obtained from 
DotAstro.org.} The ASAS-3 $V$-band light curves for the eight new RCB stars and DYPers are shown in Figure~\ref{fig:RCBlc}.

\subsection{Spectroscopy}

\begin{figure*}
\begin{center}
\includegraphics[width=120mm]{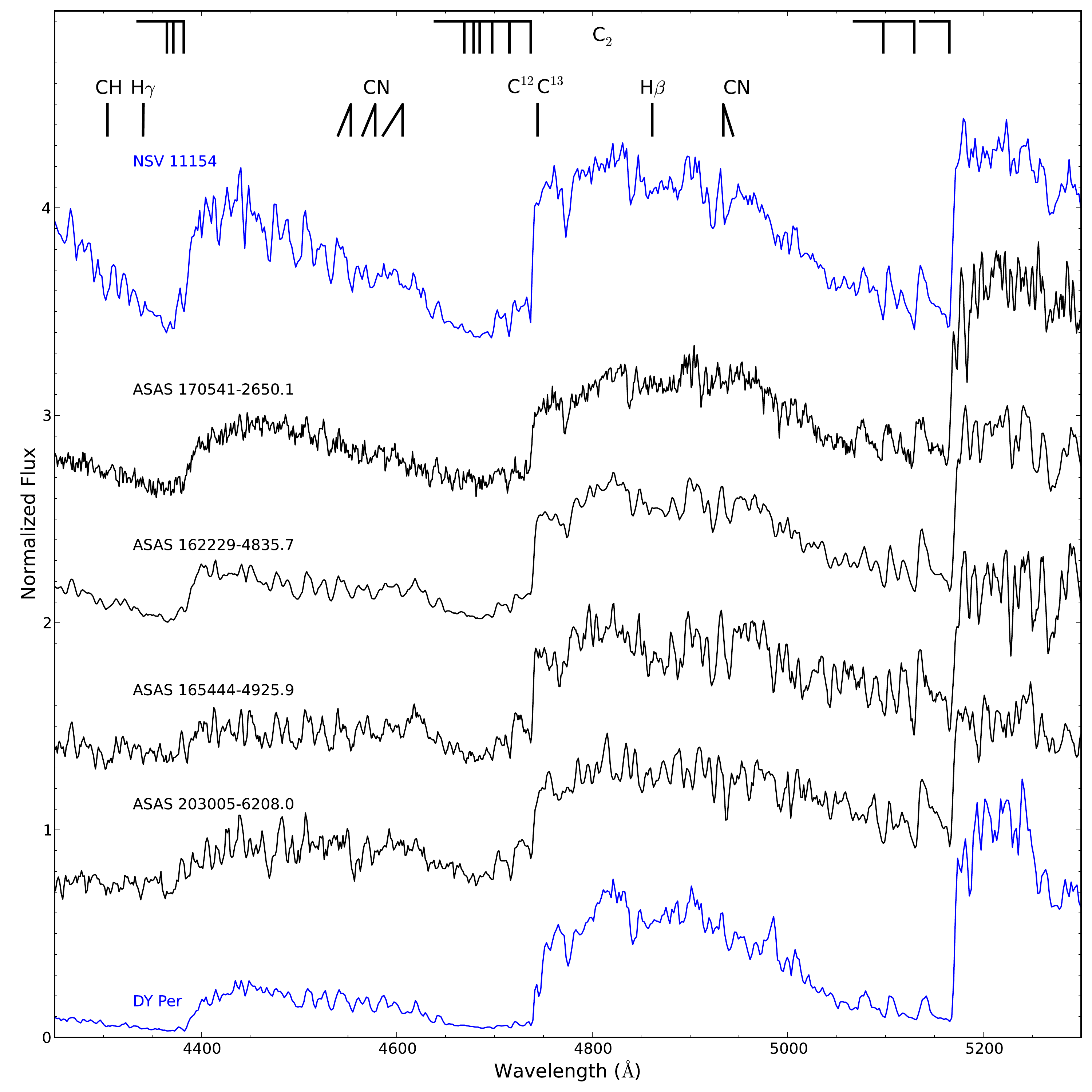}
\caption{Blue optical spectra of the new candidate RCB stars. For reference low-resolution spectra of NSV~11154, a cool RCB 
star, and DY Per obtained in early 2012 are shown in blue. The new RCB stars all show clear evidence for strong 
molecular carbon absorption and lack clear evidence for $^{13}$C as the $\lambda$4744 band of $^{12}$C$^{13}$C is 
undetected in each. There is also a lack of evidence for strong H absorption as is expected for RCB stars. 
}
\label{fig:RCBbluespec}
\end{center}
\end{figure*}

\begin{figure*}
\begin{center}
\includegraphics[width=120mm]{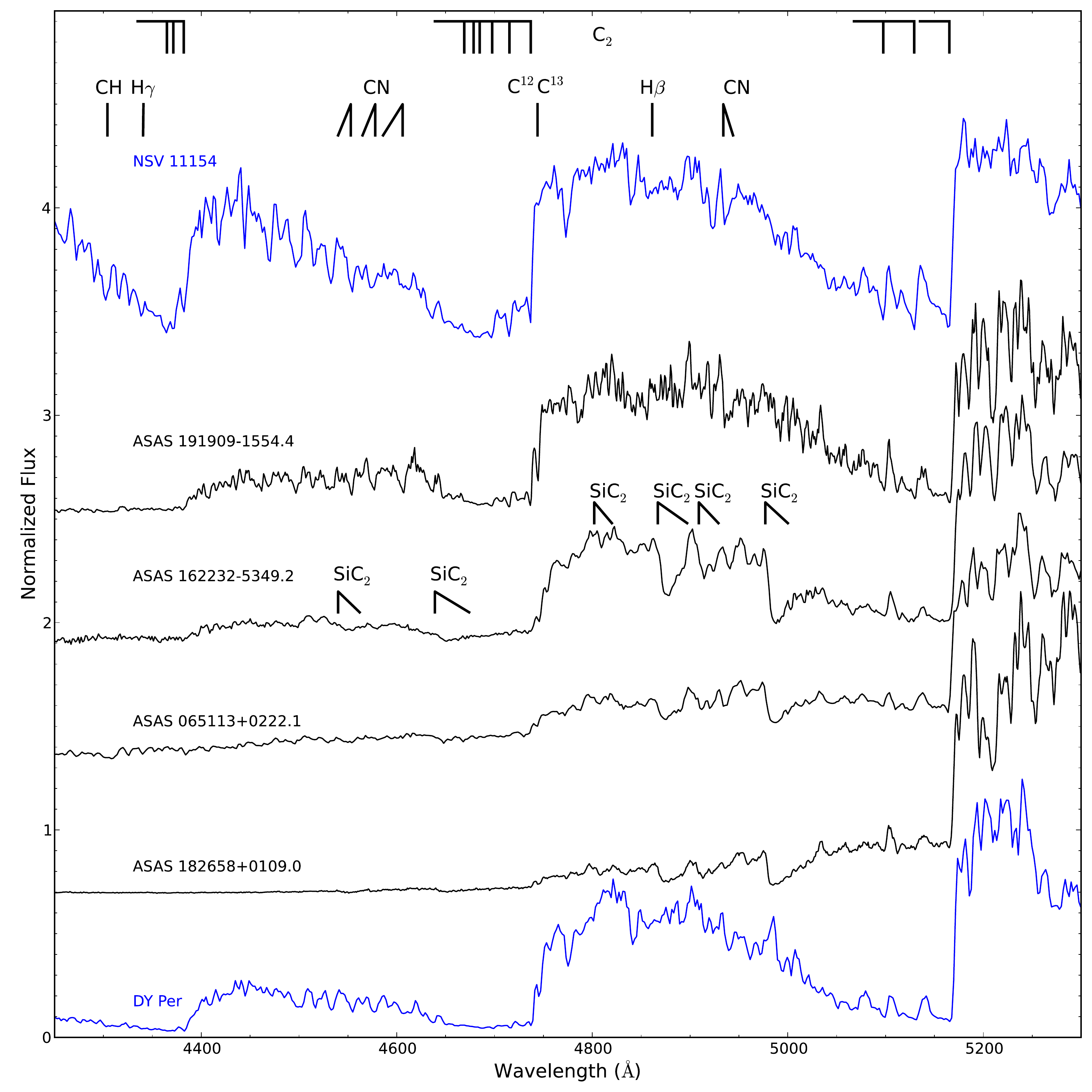}
\caption{Blue optical spectra of the four new DYPer candidates. For reference low-resolution spectra of NSV~11154, a cool RCB 
star, and DY Per obtained in early 2012 are shown in blue. The new DYPers show strong absorption from the carbon Swan bands and 
the $\lambda$4744 band of $^{12}$C$^{13}$C is clearly detected in each, similar to DY~Per. The candidates also show a lack of 
clear evidence for H absorption. Strong absorption from the Merrill-Sanford bands of SiC$_2$ is seen in three of the DYPers: 
ASAS 162232$-$5349.2, ASAS 065113+0222.1, and ASAS 182658+0109.0.}
\label{fig:DYPerbluespec}
\end{center}
\end{figure*}

Optical spectra of the candidate RCB stars were obtained between 2011 Sep.\ and 2012 May with the Kast 
spectrograph on the Lick 3-m Shane telescope on Mt.\ Hamilton, California \citep{Miller93}, the Low-Resolution 
Imaging Spectrometer (LRIS) on the 10-m Keck I telescope on Mauna Kea \citep{oke95}, and the RC Spectrograph 
on the SMARTS 1.5-m telescope at the Cerro Tololo Inter-American Observatory \citep{Subasavage10}. All spectra 
were obtained via long slit observations, and the data were reduced and calibrated using standard procedures 
(e.g., \citealt{matheson00,Silverman12}). 
On each night of observations, we obtained spectra of spectrophotometric standards to provide relative flux 
calibration for our targets. For queue-scheduled observations on the RC spectrograph, all observations in a single 
night are conducted with the slit at the same position angle. Thus, the standard stars and targets were not all observed 
at the parallactic angle, leading to an uncertain flux correction \citep{filippenko82}. We note, however, that the 
uncertainty in the flux correction does not alter any of the conclusions discussed below. A summary of our observations 
is given in Table~\ref{tab:spec}, while the blue portion of the optical spectra are shown in 
Figures~\ref{fig:RCBbluespec}--\ref{fig:DYPerbluespec}.

\begin{deluxetable*}{lccrrr} 
\tablecolumns{6}
\tabletypesize{\tiny}
\tablewidth{0pc} 
\tablecaption{Log of Spectroscopic Observations.} 
\tablehead{ 
\colhead{Name} & \colhead{UT Date} & \colhead{Instrument\tablenotemark{a}} & \colhead{Range} & \colhead{Res.} & \colhead{Exp.}  \\
\colhead{} &  \colhead{} & \colhead{} & \colhead{(\AA)} & \colhead{(\AA)} & \colhead{Time (s)} 
}
\startdata 
ASAS 060105+1654.7 & 2011-08-28.644 & LRIS & 3300--5630 & 4 & 60 \\
ASAS 060105+1654.7 & 2011-08-28.644 & LRIS & 5810--7420 & 2 & 60 \\
ASAS 170541$-$2650.1 & 2011-09-26.206 & LRIS & 3300--5630 & 4 & 60 \\
ASAS 170541$-$2650.1 & 2011-09-26.205 & LRIS & 5720--7360 & 2 & 30 \\
ASAS 191909$-$1554.4 & 2011-09-26.212 & LRIS & 3300--5630 & 4 & 5 \\
ASAS 191909$-$1554.4 & 2011-09-26.212 & LRIS & 5720--7360 & 2 & 2 \\
ASAS 162232$-$5349.2 & 2012-01-16.375 & RC & 3300--9370 & 17 & 720 \\
ASAS 095221$-$4329.7 & 2012-01-19.151 & RC & 3300--9370 & 17 & 540 \\
ASAS 065113+0222.1 & 2012-02-01.204 & Kast & 3450--9850 & 4\tablenotemark{b} & 300\tablenotemark{c}  \\
ASAS 162229$-$4835.7 & 2012-02-06.373 & RC & 3300--9370 & 17 & 540 \\
ASAS 162232$-$5349.2 & 2012-02-13.325 & RC & 3660--5440 & 4 & 2700 \\
ASAS 162229$-$4835.7 & 2012-02-15.348 & RC & 3660--5440 & 4 & 2700 \\
DY Per & 2012-02-23.107 & Kast & 3450--9850 & 4\tablenotemark{b}  & 380\tablenotemark{c} \\
ASAS 073456$-$2250.1 & 2012-03-15.225 & LRIS & 3350--10000 & 4 & 60\tablenotemark{c} \\ 
ASAS 165444$-$4925.9 & 2012-03-15.651 & LRIS & 3250--10000 & 4 & 30\tablenotemark{c} \\ 
ASAS 182658+0109.0 & 2012-03-15.654 & LRIS & 3250--10000 & 4 & 30\tablenotemark{c} \\ 
ASAS 185817$-$3543.8 & 2012-03-15.667 & LRIS & 3250--10000 & 4 & 30\tablenotemark{c} \\ 
ASAS 153214$-$2854.4 & 2012-03-15.670 & LRIS & 3250--10000 & 4 & 60\tablenotemark{c} \\ 
ASAS 203005$-$6208.0 & 2012-03-23.426 & RC & 3660--5440 & 4 & 2160 \\
NSV~11154 & 2012-04-02.524 & Kast & 3450--9850 & 4\tablenotemark{b}  & 210\tablenotemark{c} \\
ASAS 194245$-$2137.0 & 2012-04-23.513 & Kast & 3450--9850 & 4\tablenotemark{b} & 120\tablenotemark{c}\\
ASAS 180823$-$4439.8 & 2012-05-17.626 & LRIS & 5420--10000 & 4 & 16 \\
\enddata
%\tablecomments{}
\tablenotetext{a}{Kast: Kast spectrograph on Lick 3-m telescope. LRIS: low resolution imaging spectrometer on Keck-I 10-m telescope. RC: RC spectrograph on the SMARTS 1.5-m telescope.}
\tablenotetext{b}{The Kast spectrograph is a dual arm spectrograph with a resolution of $\sim$4 \AA\ on the blue side, which is relevant for the spectra shown in Figure~\ref{fig:RCBbluespec}. The typical resolution on the red side is $\sim$10 \AA.}
\tablenotetext{c}{Exposure time for the blue arm of the spectrograph. Due to the red nature of the SED the exposure time 
for the red arm was shorter than the blue.}

\label{tab:spec}
\end{deluxetable*}%

\section{Analysis}

\subsection{Spectroscopic Confirmation}

While the unique photometric behavior of RCB stars makes them readily identifiable in well sampled light 
curves taken over the course of several years, there are several examples of high-amplitude 
variables being classified as RCB stars which are later refuted by spectroscopic observations. 
Most of the misidentified candidates are either cataclysmic, symbiotic or semi-regular variables (see e.g., 
\citealt{Lawson90a, Tisserand08}). RCB stars are a subclass of the HdC stars. For an 
RCB candidate to be confirmed as a true member of the class, its spectrum must show the two prominent features 
of HdC stars: anomalously strong carbon absorption and a lack of atomic and molecular H features. 

To confirm the RCB candidates found in the ACVS, we obtained low-resolution spectra of  
the 15 candidates presented in \S~\ref{sec:cands}. Candidates observable from the northern hemisphere were observed with 
Kast and LRIS, while those only accessible from the southern hemisphere were observed with the RC spectrograph. 
For some of the southern hemisphere targets very low resolution spectra were obtained first to confirm the presence of 
C$_2$ before slightly higher resolution observations were obtained (see Table~\ref{tab:spec}).

We searched the spectra for the presence of strong carbon features, primarily C$_2$ and CN, and a lack of Balmer 
absorption to confirm the RCB classification for the ACVS candidates. We find these characteristics in eight of the 
spectroscopically observed stars (see Figures~\ref{fig:RCBbluespec}--\ref{fig:DYPerbluespec}), which we consider 
good RCB and DYPer candidates as summarized in Table~\ref{tab:observations}. The remaining candidates were rejected as 
possible RCB stars based on their spectra, which typically showed strong TiO and VO absorption or clear evidence for 
H. The properties of the rejected candidates are summarized in Table~\ref{tab:rejects}.
In the remainder of this paper we no longer consider these stars candidates and restrict our discussion to 
the eight good candidates listed in Table~\ref{tab:observations}. 

In addition to the hallmark traits of overabundant carbon and a lack of Balmer absorption, RCB stars show a number 
of other unique spectroscopic characteristics. In particular, they show a very high ratio of $^{12}$C/$^{13}$C 
and no evidence for $G$ band absorption. To search for the presence of $^{13}$C, we examined the spectra for 
the $\lambda$4744 band head of $^{12}$C$^{13}$C, which is typically very weak or absent in 
the spectra of RCB stars. We find evidence for $^{12}$C$^{13}$C in ASAS 191909$-$1554.4, ASAS 162232$-$5349.2, 
ASAS 065113+0222.1, and ASAS 182658+0109.0 while ASAS 162232$-$5349.2 shows possible evidence for the $^{13}$C$^{13}$C band 
at $\lambda$4752. The presence of $^{13}$C suggests that these four stars are likely DYPers. 
We consider these four stars closer analogs to DY Per and the DYPers found in the LMC 
and SMC \citep{Alcock01,Tisserand09} than they are to classical RCB stars. One of the DYPers,  
ASAS 065113+0222.1, shows weak evidence for CH $\lambda$4300 ($G$ band) absorption and possible evidence for 
H$\gamma$, which is sometimes seen in the spectra of DYPers. We note that the signal-to-noise ratio (S/N) of all our 
spectra in the range between $\sim$4300--4350 is relatively low, making definitive statements about the 
presence or lack of both CH and H$\gamma$ challenging. Finally, we note that we see evidence for the 
Merrill-Sanford bands of SiC$_2$ in three of our candidates: ASAS 162232$-$5349.2, ASAS 065113+0222.1, and ASAS 182658+0109.0. 
To our knowledge this is the first identification of SiC$_2$ in a DYPer spectrum, though the presence of 
this molecule should not come as a surprise as RCB stars are both C and Si rich \citep{Clayton96}.

\begin{deluxetable*}{lrrrrcccl} 
\tablecolumns{9}
\tabletypesize{\tiny}
\tablewidth{0pc} 
\tablecaption{Observational Properties of New RCB Stars and DYPers.}
\tablehead{ 
\colhead{Name} & \colhead{$V_{\rm max}$\tablenotemark{a}} & \colhead{$\Delta$mag} & \colhead{$\Delta$t} & \colhead{$dm/dt$} & \colhead{$^{13}$C} & \colhead{H/CH} & \colhead{pulsations} & \colhead{RCB/}  \\
\colhead{} &  \colhead{(mag)} & \colhead{} & \colhead{(d)} & \colhead{(mag day$^{-1}$)} & \colhead{} & \colhead{} & \colhead{} & \colhead{DYPer} 
}
\startdata 
ASAS 170541$-$2650.1 & 11.9 & 1.2 & 31 & 0.04 & Weak 4744? & None & N\tablenotemark{b} & RCB \\
ASAS 162229$-$4835.7 & 10.8 & 2.8 & 83 & 0.03 & None & Weak H$\gamma$?, Weak CH & Y & RCB \\
ASAS 165444$-$4925.9 & 11.8 & $>$1.6 & $>$48 & 0.03 & None & weak H$\gamma$?, H$\beta$? & Y & RCB \\
ASAS 203005$-$6208.0 & 13.2 & $>$1.4 & $>$30 & 0.05 & None & H$\gamma$?, H$\beta$? -- blends & Y & RCB \\
ASAS 191909$-$1554.4 &  6.9 & 1.0 & 20 & 0.05 & Y & None & Y & DYPer \\
ASAS 162232$-$5349.2 & 11.5 & 1.7 & $<$256 & $>$0.007 & Y & None & Y & DYPer \\
ASAS 065113+0222.1 & 12.4 & 1.0 & $<$140 & $>$0.007 & Y & weak H$\gamma$?, Weak CH & Y & DYPer \\
ASAS 182658+0109.0 & 12.1 & 1.6 & 960 & 0.002 & Y & None & Y & DYPer \\
\enddata
%\tablecomments{}
\tablenotetext{a}{Observed quantity, not corrected for Galactic reddening.}
\tablenotetext{b}{The period of ASAS observations covers very little time around maximum light, and as a result there is a relatively short period of data suitable for searching for pulsations. See \S~\ref{sec:pulsations}.}

\label{tab:observations}
\end{deluxetable*}%

\begin{deluxetable*}{lrl} 
\tablecolumns{3}
\tabletypesize{\tiny}
\tablewidth{0pc} 
\tablecaption{Rejected RCB Candidates.} 
\tablehead{ 
\colhead{Name} & \colhead{$V_{\rm max}$\tablenotemark{a}} & \colhead{Remarks} \\
\colhead{} &  \colhead{(mag)} & \colhead{} 
}
\startdata 
ASAS 060105+1654.7 & 12.3 & No C$_2$; strong H, $G$ band \\
ASAS 073456$-$2250.1 & 12.8 & C$_2$; strong H emission \\
ASAS 095221$-$4329.7 & 10.6 & Strong TiO, VO; H? \\
ASAS 153214$-$2854.4 & 12.3 & No C$_2$; H$\alpha$ emission, strong $G$ band; SRPV\tablenotemark{b} \\
ASAS 180823$-$4439.8 & 12.1 & Strong TiO, VO \\
ASAS 185817$-$3543.8 & 10.9 & Strong TiO, VO \\
ASAS 194245$-$2137.0 & 12.5 & Strong TiO, VO \\
\enddata
%\tablecomments{}
\tablenotetext{a}{Observed quantity, not corrected for Galactic reddening.}
\tablenotetext{b}{Identified as a semi-regular periodic variable in the General Catalog of Variable Stars \citep{Samus08}.}

\label{tab:rejects}
\end{deluxetable*}%

\subsection{Photometric Behavior}

In addition to spectral differences, RCB stars and DYPers show some dissimilarities in their photometric 
evolution as well. The first order behavior is the same: both show deep, irregular declines in 
their light curves which can take anywhere from a few months to a several years to recover to maximum 
brightness. Beyond that generic behavior, however, the shape of the decline tends to differ: RCB stars show fast declines 
with slow recoveries whereas DYPers tend to show a more symmetric decline and recovery.

The photometric properties of our candidates, including decline rates for the most prominent and well sampled 
declines, are summarized in Table~\ref{tab:observations}. 
As previously noted in the caption of Figure~\ref{fig:RCBfinders}, 
the full amplitude of the variations of these stars are likely underestimated due to the 
large PSF on ASAS images. This means that the decline rates should be treated as lower limits, since the 
true brightness of the star may be below that measured in a large aperture. Nevertheless, the decline rates for the 
four RCB stars are relatively fast and consistent with those given in \citet{Tisserand09} for RCB stars in 
the MCs, $\sim$0.04 mag day$^{-1}$. The most telling feature of the light curves is the shape of 
the declines, however. For the four spectroscopic RCB stars, ASAS 170541$-$2650.1, ASAS 162229$-$4835.7, 
ASAS 165444$-$4925.9, and ASAS 203005$-$6208.0 the declines 
are very rapid. While we do not detect ASAS 165444$-$4925.9 after its sharp decline on TJD$\approx$5000, the other three 
show slow asymmetric recoveries to maximum light. The four spectroscopic DYPers generally show a slower decline 
with a roughly symmetric recovery, though we note that the full recoveries of ASAS 065113+0222.1 and ASAS 162232$-$5349.2 
are not observed. 

\subsection{Pulsations}\label{sec:pulsations}

All RCB stars are variable near maximum light, with most and possibly all of the variations thought to be due to pulsation 
\citep{Lawson90b,Clayton96}. Typical periods are $\sim$40--100 days, and the amplitudes are a few tenths of a magnitude. 
The pulsational properties of DYPers are not as well constrained, because the sample is both small 
and only recently identified. Each of the four DYPers identified in \citet{Alcock01} shows evidence for 
periodic variability near maximum light, with typical periods of $\sim$100--200 days.

To search for the presence of pulsations in our candidate RCB stars, we use a generalized Lomb--Scargle 
periodogram (\citealt{Lomb76,Scargle82,Zechmeister09}) to analyze each star (see 
\citealt{richards11} for more details on our Lomb--Scargle periodogram implementation). Our analysis only 
examines data that are well separated from decline phases, and we focus on the portions of light curves where the 
secular trend is slowly changing relative to the periodic variability. For each star we simultaneously fit for the 
harmonic plus linear or quadratic long-term trend in the data. The frequency that produces the largest peak in the 
periodogram, after masking out the 1 day alias, is adopted as the best-fit period. 

\begin{figure}
\begin{center}
\includegraphics[width=90mm]{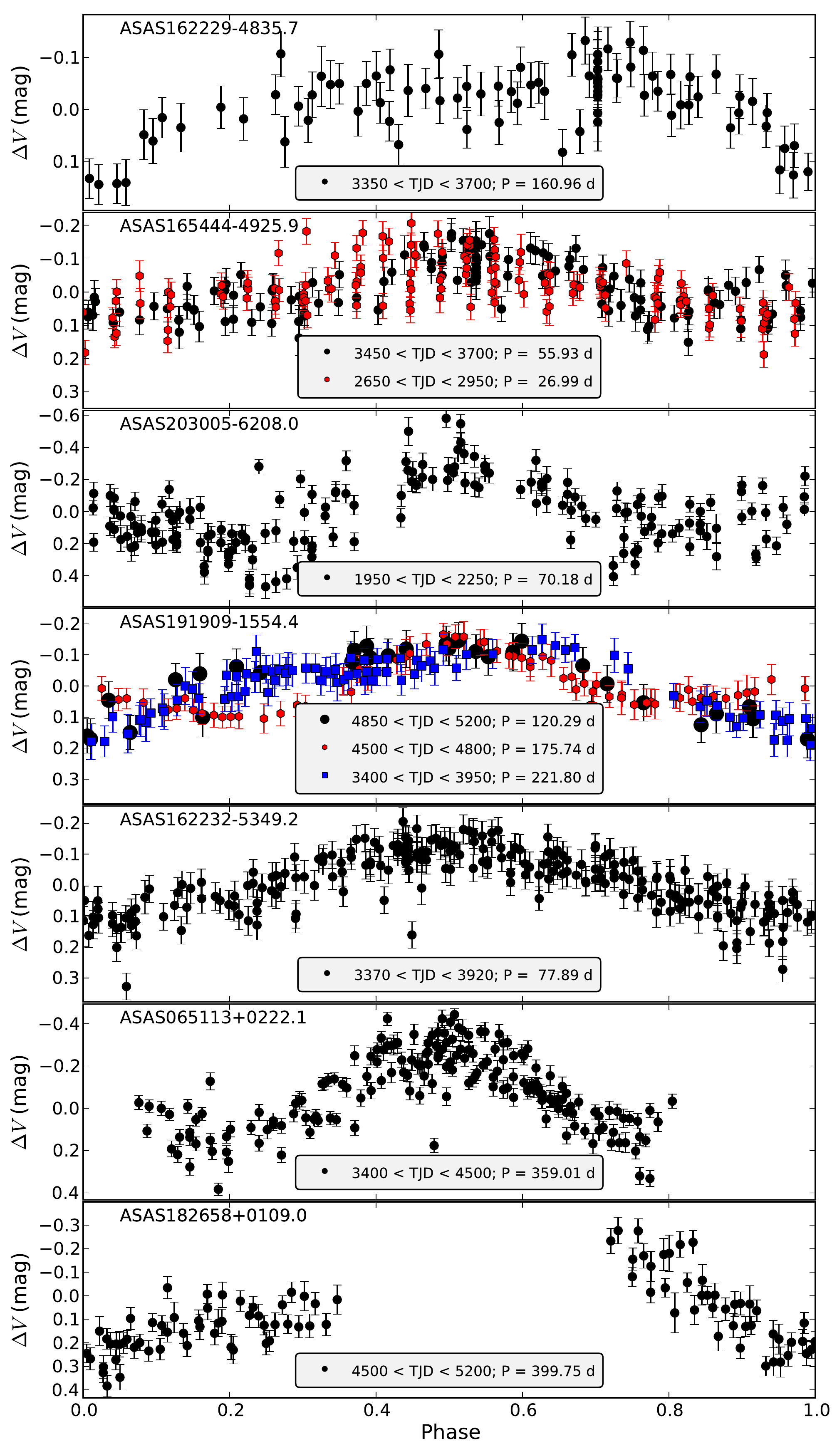}
\caption{Folded light curves showing evidence for periodic variability near maximum light in the new RCB stars and DYPer 
candidates. The folded light curves only display a portion of the ASAS observations as indicated in the legend for each 
source. ASAS 165444$-$4925.9 and ASAS 191909$-$1554.4 show evidence for multiple dominant periods, which are shown 
with blue and red points as indicated in their respective legends.}
\label{fig:RCBpulse}
\end{center}
\end{figure}

We find evidence for periodicity in the light curves of each star, except 
for ASAS 162229$-$4835.7. For most of the observing window ASAS 162229$-$4835.7 was in or near a decline, 
and we predict that additional observations of ASAS 162229$-$4835.7, be they historical or in the future, will show periodic 
variability near maximum light. The trend-removed, phase-folded light curves of the remaining seven stars are shown in 
Figure~\ref{fig:RCBpulse}. Insets in each panel list the range of dates that were included in the Lomb-Scargle analysis, 
as well as the best fit period for the data. 

Some RCB stars are known to have more than a single dominant period (see \citealt{Clayton96} and references therein). 
We find evidence for multiple periods in ASAS 191909$-$1554.4, with periods of 120, 175, and 221 days that appear to change 
every $\sim$1--2 years. Evidence for multiple periods also appears to be present in ASAS 165444$-$4925.9. 
The best fit periods in this case are 27 and 56 days, which differ by roughly a factor of two. The longer period may in this 
case simply be a harmonic of the shorter period. Finally, we note that the best fit period for ASAS 162232$-$5349.2, 359 days, 
is very close to one year, and it is possible that the data are beating against the yearly observation cycle. The folded 
light curve appears to traverse a full cycle over $\sim$half the full phase cycle. The slight upturn in the 
folded data around phase 0.15 suggests that the true period is likely $\sim$180 days, half the best fit period.

\subsection{Spectral Energy Distributions}\label{sec:SED}

All RCB stars are known to have an infrared (IR) excess due to the 
presence of circumstellar dust \citep{Feast97a,Clayton96}, and all the known DYPers in the MCs also show 
evidence for excess IR emission \citep{Tisserand09}. To check for a similar excess in the new ACVS RCB stars and DYPers,  
broadband spectral energy distributions (SEDs) were constructed 
with catalog data obtained from USNO-B1 \citep{monet03}, the Two Micron All Sky Survey (2MASS; 
\citealt{skrutskie-2mass}), the {\it Wide-Field Infrared Survey Explorer} ({\it WISE}, \citealt{Wright10}), 
{\it AKARI} \citep{murakami07}, the {\it Mid-course Space 
Experiment} ({\it MSX}; \citealt{mill94}), and the {\it Infrared Astronomical Satellite} ({\it IRAS}; 
\citealt{Neugebauer84}).\footnote{Catalog data for each of these surveys can be found at: 
{\tt http://irsa.ipac.caltech.edu/}} The USNO-B1 catalog contains measurements made on digital scans 
of photographic plates corresponding roughly to the $B$, $R$, and $I$ bands. Repeated $B$ and $R$ plates were taken 
typically more than a decade apart. To convert the five separate USNO-B1 magnitude 
measurements to the standard $g'r'i'$ system of the Sloan Digital Sky Survey (SDSS; \citealt{fukugita96}), we invert the 
filter transformations from (\citealt{monet03}; see also 
\citealt{Sesar06}). The two measurements each for the $g'$ and the $r'$ band are then averaged to get the reported SDSS 
$g'$ and $r'$ magnitudes, unless the two measurements differ by $>$ 1 mag, in which case it is assumed that the fainter 
observation occurred during a fading episode of the star. Then the final adopted SDSS magnitude is that of the brighter 
measurement. There is a large scatter in the transformations from USNO-B1 to SDSS \citep{monet03,Sesar06}, which,
leads us to adopt a conservative 1-$\sigma$ uncertainty of 40\% in flux density on each of the transformed SDSS 
flux measurements. The 
2MASS magnitude measurements are converted to fluxes via the calibration of \citet{cohen2003} and the {\it WISE} magnitudes 
are converted to fluxes via the calibration in \citet{Cutri11a}. The remaining catalogs provide flux density measurements 
in Jy rather than using the Vega magnitude system.

\begin{figure}
\begin{center}
\includegraphics[width=90mm]{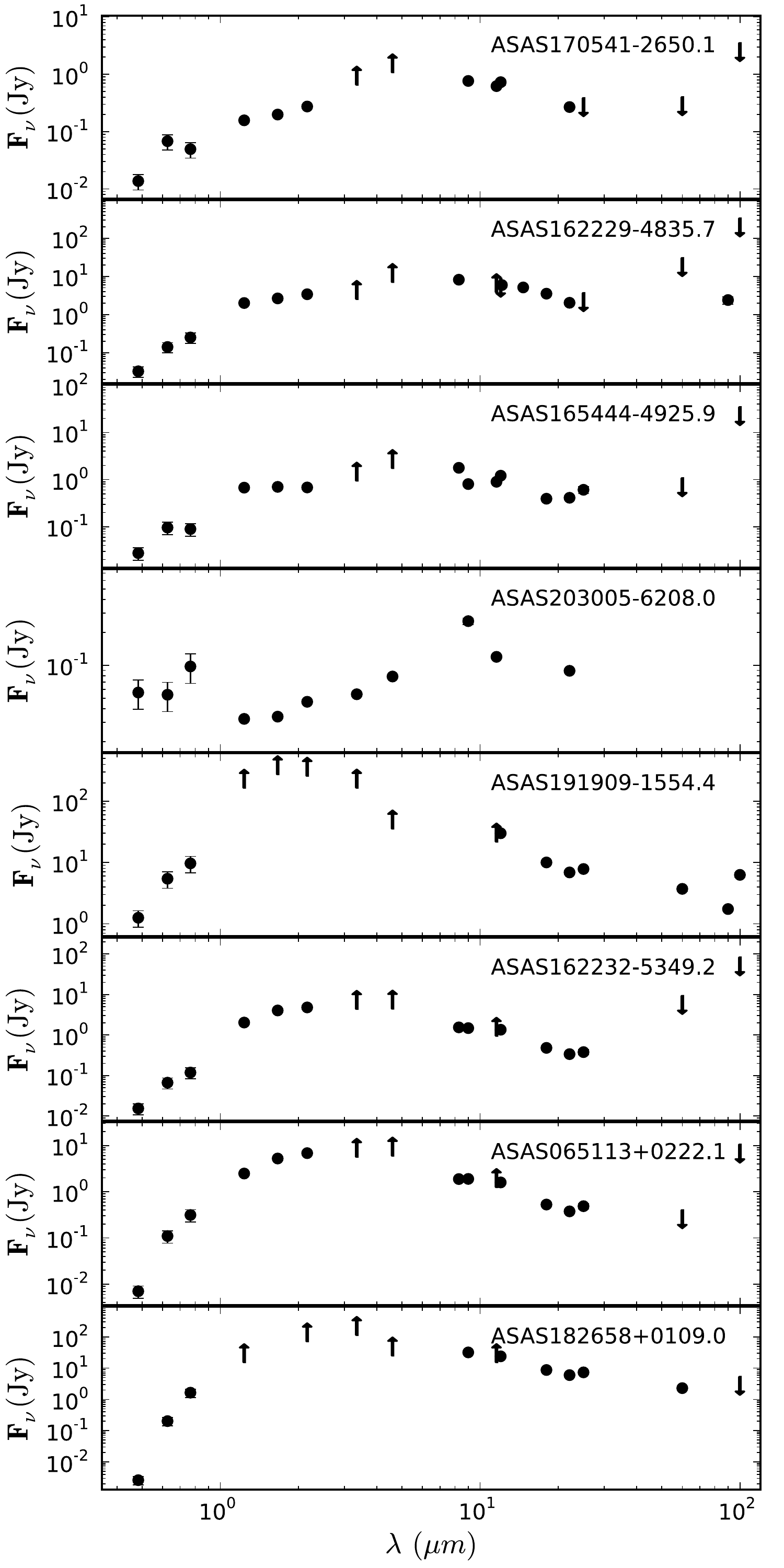}
\caption{Optical through mid-infrared spectral energy distributions for newly discovered RCB stars and DYPers. 
The observations in the various passbands were not taken simultaneously, thus in some cases the lack of a smooth spectrum 
is likely the result of intrinsic variability. The reddening towards each star is uncertain.}
\label{fig:RCBSED}
\end{center}
\end{figure}

The full SEDs extending from the optical to the mid-infrared for each of the new RCB stars and DYPers are shown in 
Figure~\ref{fig:RCBSED}. All of the candidates but ASAS 203005$-$6208.0 saturate the W1 (3.5 $\mu$m) and W2 (4.6 $\mu$m) 
bands of {\it WISE}, while of those all but ASAS 170541$-$2650.1 and ASAS 165444$-$4925.9 saturate W3 (11.6 $\mu$m) as well. 
ASAS 191909$-$1554.4 and ASAS 182658+0109.0 saturate all three of the 2MASS filters and are the only candidates 
detected at either 60 and/or 100 $\mu$m by {\it IRAS}. ASAS 162229$-$4835.7 and ASAS 191909$-$1554.4 were 
the only candidates detected at 90 $\mu$m by {\it AKARI}. 

Using RCB stars and DYPers in the MCs, \citet{Tisserand09} find that the RCB stars typically have SEDs with two distinct 
peaks, whereas DYPers typically have a single peak. It is argued in \citeauthor{Tisserand09} that the SEDs of both can 
be understood as emission from a stellar photosphere and surrounding dust shell; the cooler photospheric temperatures 
of DYPers are less distinct relative to the dust emission leading to a single broad peak rather than two.
We caution that the reddening toward each of the new Galactic candidates is 
unknown, which makes a detailed analysis of their SEDs challenging. Furthermore, the observations were not taken 
simultaneously in each of the various bandpasses. Nevertheless, a few interesting trends can be gleaned from the data. 
The four stars that spectroscopically resemble RCB stars, ASAS 170541$-$2650.1, ASAS 162229$-$4835.7, ASAS 165444$-$4925.9, 
and ASAS 203005$-$6208.0 all show clear evidence for a mid-infrared excess relative to their optical brightness. The 
peak of emission from ASAS 191909$-$1554.4 and ASAS 182658+0109.0 is not well constrained 
because they saturate the detectors between 1 and 6 $\mu$m, yet interestingly both show evidence for an infrared excess 
redward of 50 $\mu$m. This suggests that there might be some very cool ($T <$ 100 K) dust 
in the circumstellar environment of these stars, which is observed in some of the bright, nearby RCB stars. For instance, 
{\it Spitzer} and {\it Herschel} observations of R CrB show evidence for a 
large, $\sim$4 pc, cool, $T \sim 25$ K, and diffuse shell of gas that is detected in the far-IR \citep{Clayton11}. 
ASAS 162232$-$5349.2 and ASAS 065113+0222.1 show evidence for a single broad 
peak in their SED occurring around $\sim$2 $\mu$m, which is similar to the SEDs of the DYPers observed in the MCs. 

\subsection{Near-infrared Variability}\label{sec:NIRvar}

An overlap in the survey fields between 2MASS and the DEep Near Infrared Souther Sky Survey (DENIS; \citealt{Epchtein94}) 
allows measurements of the NIR variability of four of the newly discovered RCB stars and DYPers. 
Photometric measurements from 2MASS and the two epochs of DENIS observations for these 
stars are summarized in Table~\ref{tab:IR_var}. Unfortunately the 2MASS and DENIS observations proceeded the  
ASAS monitoring, and so we cannot provide contextual information such as the state of the star (near maximum, on decline, 
during deep minimum, etc.) at the time of the NIR observations. To within a few tenths of a magnitude, ASAS 162229$-$4835.7 is 
not variable between the 2MASS and DENIS observations. On timescales of a few weeks to months both ASAS 162232$-$5349.2 and 
ASAS 170541$-$2650.1 show evidence for variations $\gtrsim$ 1 mag in the NIR. Similar variations have been observed for 
several of the RCB stars and DYPers in the MCs (e.g., \citealt{Tisserand04,Tisserand09}). The largest variations 
were observed in ASAS 203005$-$6208.0, which changed by $\sim$4 mag in the $J$ band during the $\sim$4 yr between the DENIS 
and 2MASS observations. ASAS 203005$-$6208.0 also shows a large variation between the DENIS $I$ band measurement 
and the $I$-band measurement from USNO-B1, with $\Delta{m} \approx 7$ mag. This star is clearly a large-amplitude variable, 
which likely explains its unusual SED. In the DENIS observations, which provide simultaneous optical and NIR measurements, 
ASAS 203005$-$6208.0 is always fainter in the optical, suggesting that the unusual shape to its SED (see 
Figure~\ref{fig:RCBSED}) is the result of non-coeval observations.

\begin{deluxetable*}{lrrrrrrrr} 
\tablecolumns{9}
\tabletypesize{\tiny}
\tablewidth{0pc} 
\tablecaption{2MASS and DENIS NIR Measurements.} 
\tablehead{ 
\colhead{Name} & \colhead{Epoch$_{\rm 2MASS}$\tablenotemark{a}} & \colhead{$J_{\rm 2MASS}$\tablenotemark{a}} & \colhead{$H_{\rm 2MASS}$\tablenotemark{a}} & \colhead{$K_{s{\rm 2MASS}}$\tablenotemark{a}} & \colhead{Epoch$_{\rm DENIS}$\tablenotemark{b}} & \colhead{$I_{\rm DENIS}$\tablenotemark{b}} & \colhead{$J_{\rm DENIS}$\tablenotemark{b}} & \colhead{$K_{s{\rm DENIS}}$\tablenotemark{b}} \\
\colhead{} & \colhead{(JD)} & \colhead{(mag)} & \colhead{(mag)} & \colhead{(mag)} & \colhead{(JD)} & \colhead{(mag)} & \colhead{(mag)} & \colhead{(mag)}
}
\startdata 
ASAS 170541$-$2650.1 & 2451004.658 & 10.01 & 9.28 & 8.46 & 2451730.639 & 11.09 & 10.18 & 9.13 \\
 												& & & & & 2451749.592 & 9.94 & 8.91 & 7.96 \\
ASAS 162229$-$4835.7 & 2451347.541 & 7.26 & 6.47 & 5.73 & 2451387.540 & 9.10 & 6.90 & 5.40 \\
 												& & & & & 2451395.492 & 8.98 & 7.13 & 5.45 \\
ASAS 162232$-$5349.2 & 2451347.538 & 7.24 & 6.01 & 5.36 & 2451387.532 & 9.12 & 6.59 & 4.40 \\
 												& & & & & 2451395.485 & 9.20 & 6.30 & 4.28 \\
ASAS 203005$-$6208.0 & 2451701.878 & 11.73 & 11.20 & 10.40 & 2450267.768 & 17.62 & 15.66 & 12.06 \\
 												& & & & & 2451003.746 & 15.82 & 14.37 & 11.89 \\
\enddata
\tablecomments{}
\tablenotetext{a}{Catalog measurement from the 2MASS point source catalog \citep{cutri03}.}
\tablenotetext{b}{Catalog measurement from the DENIS point source catalog \citep{Epchtein99}.}

\label{tab:IR_var}
\end{deluxetable*}%

\section{Individual Stars}

We discuss the individual stars and whether they should be considered RCB stars or DYPers below. We also use SIMBAD to 
identify alternate names for these stars and previous studies in the literature (see also Table~\ref{tab:coords}).  

\subsection{ASAS 170541$-$2650.1 (GV Oph)}

This star was first identified as a variable source on Harvard photographic plates with the name Harvard Variable 4368, 
and was cataloged as a likely long period variable based on the large amplitude of variations from 13.9 mag to below the 
photographic limit of $\sim$16.5 mag \citep{Swope28}. It was later named GV Oph in the General Catalog of Variable Stars 
(GCVS; \citealt{Kukarkin71,Samus08}) as a variable of unknown type with rapid variations. The light curve, spectrum, and SED 
of this star are consistent with it being an RCB star.

\subsection{ASAS 162229$-$4835.7 (IO Nor)}  

This star is listed as a Mira variable in the GCVS with the name IO Nor. In \citet{Clarke05} it is identified as a 
star with an IR excess based on {\it MSX} observations. On the basis of its NIR and mid-IR colors, it is identified as a 
RCB candidate in \citet{Tisserand12}, and considered a likely RCB star on the basis of its ASAS light curve.\footnote{In 
\citet{Tisserand12} two additional stars with IR colors consistent with RCB stars, V653 Sco and V581 CrA, are 
identified as highly likely RCB stars on the basis of their ASAS light curves. V581 CrA is not included in ACVS and therefore 
is not included in the MACC. V653 Sco is listed in the GCVS as a Mira variable and classified as a semi-regular periodic 
variable in the MACC, $P({\rm SRPV}) = 0.55$. It has $P({\rm RCB}) = 0.012$ and $R({\rm RCB}) = 2609$. The light curve 
is somewhat similar to ES Aql, in that it fades below the ASAS 
detection limits and it is highly active during the $\approx$8 yr it was observed, meaning it that folds decently well on a period 
of $\sim$450 days. A spectrum will be needed to disambiguate between an RCB and long period variable classification for V653 
Sco.} We independently identified IO Nor as a likely RCB star on the basis of its light curve (in the MACC it is the most 
likely RCB in ACVS), and our spectrum confirms that it is a genuine RCB star. The previous 
classification as a Mira variable is likely based on the late spectral type and large amplitude of variability, but 
Figure~\ref{fig:RCBlc} clearly shows that ASAS 162229$-$4835.7 is not a long period variable.

\subsection{ASAS 165444$-$4925.9 (C* 2377)}

The variability of this star has not been cataloged to date, and it is listed in the CGCS as C* 2377 \citep{Alksnis01}.
The spectrum, SED, and pulsations exhibited by this star are consistent with RCB stars. There may be evidence for weak CH 
absorption, though we caution that the S/N is low near $\sim$4300 \AA. The light curve shows a sharp decline, similar to RCB 
stars, but the recovery is not observed. Nevertheless, the evidence points to ASAS 165444$-$4925.9 being a new member 
of the RCB class.

\subsection{ASAS 203005$-$6208.0 (NSV~13098)}

This star was first identified as variable by \citet{Luyten32} with a maximum brightness of 14 mag and a minimum $>$ 18 mag. 
\citeauthor{Luyten32} assigned it the name AN~141.1932, and it was later cataloged as a possible variable star in the GCVS as 
NSV~13098. The light curve, spectrum, and SED are all consistent with an RCB classification. Higher resolution and S/N 
spectra are needed to confirm if H absorption is present, though we note that some RCB stars do show evidence of H in their 
spectra (e.g., V854 Cen; \citealt{Kilkenny89}), leading us to conclude that ASAS 203005$-$6208.0 is an RCB star. 

\subsection{ASAS 191909$-$1554.4 (V1942 Sgr)}

This star is listed as a slow irregular variable of late spectral type with the name V1942 Sgr in the GCVS. It is the 
brightest star among our candidates, and as such it is one of the best studied carbon stars to date. According to SIMBAD 
ASAS 191909$-$1554.4 is discussed in more than 50 papers in the literature. In the CGCS it is listed as C* 2721. 
ASAS 191909$-$1554.4 is detected by {\it Hipparcos} (HIP 94940) and has a measured parallax of 2.52$\pm$0.82 mas 
\citep{Perryman97}. This corresponds to a distance $d = 397 \pm 115$ pc and a distance modulus $\mu \approx$ 8 
mag. ASAS 191909$-$1554.4 is one of the few Galactic carbon stars with a measured parallax, and it is important 
for constraining the luminosity function of carbon stars \citep{Wallerstein98}. In their spectral atlas of carbon stars, 
\citet{Barnbaum96} identify ASAS 191909$-$1554.4 as having a spectral type of N5+ C$_2$5.5. Relative abundance measurements 
by \citet{Abia97} show that $^{12}$C/$^{13}$C$~=~30$, which is low relative to classical RCB stars. The proximity of 
ASAS 191909$-$1554.4 allows its circumstellar dust shell to be resolved in {\it IRAS} images (e.g., \citealt{Young93}), and 
\citet{Egan91} use ASAS 191909$-$1554.4 and other carbon stars with resolved dust shells and 100 $\mu$m excess to 
statistically argue that each of these stars must be surrounded by two dust shells, one that is old, $\sim$10$^4$ yr, 
and the other that is produced by a current episode of mass loss. Recent \ion{H}{1} observations by 
\citet{Libert10} have shown evidence for the presence of H in the circumstellar shell of ASAS 191909$-$1554.4. The shallow, 
symmetric fade of the light curve, along with the N type carbon star spectrum and the presence of $^{13}$C in the spectrum, leads
us to conclude that ASAS 191909$-$1554.4 is a DYPer. This is the only candidate within our sample for which we can 
measure the absolute magnitude, since we have the {\it Hipparcos} parallax measurement. Adopting a maximum light brightness of 
$V_{\rm max} =$ 6.8 mag, we find that ASAS 191909$-$1554.4 has $M_V \approx -1.2$ mag. This is roughly 0.4 mag fainter than 
the faintest DYPers in the MCs \citep{Tisserand09}, suggesting that either the luminosity function extends fainter than that 
observed in the MCs or there is unaccounted for dust extinction toward ASAS 191909$-$1554.4.

\subsection{ASAS 162232$-$5349.2 (C* 2322)}

The variability of this star has not been cataloged to date, and it is listed in the CGCS as C* 2322. 
The relatively slow, symmetric decline and recovery in the light curve of ASAS 162232$-$5349.2 lead us to classify it as a 
DYPer variable. The presence of $^{13}$C in the spectrum and the single peak in the SED support this classification. 

\subsection{ASAS 065113+0222.1 (C* 596)}

The variability of this star has not been cataloged to date, and it is listed in the CGCS as C* 596. The presence of 
$^{13}$C in the spectrum and the single peak in the SED lead us to classify ASAS 065113+0222.1 as a DYPer variable. 

\subsection{ASAS 182658+0109.0 (C* 2586)}

The variability of this star has not been cataloged to date, and it is listed in the CGCS as C* 2586. Based on weekly 
averages of {\it DIRBE} NIR observations taken over 3.6 yr, \citet{Price10} list ASAS 182658+0109.0 as a non-variable 
source. Low resolution spectra taken with {\it IRAS} show 11 $\mu$m SiC dust emission, which typically indicates 
significant mass loss from a carbon star \citep{Kwok97}. The light curve shows a $\sim$5--6 yr symmetric decline, and there 
is clear evidence for $^{13}$C in the spectrum. The {\it IRAS} 
detection at 60 $\mu$m shows a clear IR excess relative to a single temperature blackbody. While there is no evidence for 
H$\alpha$, the S/N in our spectrum is low blueward of $\sim$4700 \AA. We consider ASAS 182658+0109.0 a likely 
DYPer, though higher S/N spectra are required for a detailed abundance analysis to confirm this classification.

\section{Discussion}

\subsection{New Candidates from an Expanded Training Set}\label{sec:aug_train}

\begin{deluxetable*}{lllrrcccc}
\tablecolumns{9}
\tabletypesize{\tiny}
%\rotate
\tablewidth{0pc} 
\tablecaption{New RCB/DYPer candidates using an augmented training set.} 
\tablehead{ 
\colhead{Name} & \colhead{Other ID} & \colhead{DotAstro\tablenotemark{a}} & \colhead{$\alpha_{\rm J2000.0}$\tablenotemark{b}} & \colhead{$\delta_{\rm J2000.0}$\tablenotemark{b}} & \colhead{CGCS\tablenotemark{c}} & \colhead{$P({\rm RCB})$\tablenotemark{d}} & \colhead{R({\rm RCB})\tablenotemark{e}} & Remarks \\
\colhead{} &  \colhead{} & \colhead{ID} & \colhead{(hh mm ss.ss)} & \colhead{(dd mm ss.s)} & \colhead{ID} & \colhead{new} & \colhead{new} & \colhead{}
} 
\startdata 
ASAS 053302+1808.0 & IRAS 05301$+$1805 & 219583 & 05 33 01.72 & 18 07 59.0 & 980 & 0.339 & 380 & 1 \\
ASAS 081121$-$3734.9 & C* 1086 & 227950 & 08 11 21.39 & $-$37 34 54.2 & 2106 & 0.145 & 492 & 1 \\
ASAS 125245$-$5441.6 & ... & 237449 & 12 52 44.92 & $-$54 41 37.5 & ... & 0.309 & 394 & 2 \\
ASAS 160033$-$2726.3 & 1RXS J160033.8$-$272614? & 243486 & 16 00 33.16 & $-$27 26 18.5 & ... & 0.142 & 498 & 2,3 \\
ASAS 175226$-$3411.5 & IRAS 17491$-$3410 & 249729 & 17 52 25.50	& $-$34 11 28.2 & ... & 0.166 & 460 & \\
ASAS 200531+0427.2 & V902 Aql & 259768 & 20 05 30.83 & 04 27 12.8 & ... & 0.382 & 353 & 2,4 \\
\enddata
\tablecomments{Remarks: {\it 1}. Known carbon star, shows evidence for multiple declines that may be periodic. Semi-regular variable? {\it 2}. Shows evidence for a single partially resolved decline or recovery. {\it 3}. This star is $\sim$9\arcsec\ from the cataloged X-ray source 1RXS J160033.8$-$272614. The possible association with an X-ray counterpart suggests that it may be a Be star. {\it 4}. This star is listed as having an M spectral type in the GCVS. We were unable to find a published reference listing this spectral type and suggest a new spectrum be taken to determine the spectral type.}
\tablenotetext{a}{ID from the MACC.}
\tablenotetext{b}{Reported coordinates from the Two Micron All Sky Survey point source catalog \citep{cutri03}.}
\tablenotetext{c}{ID from the General Catalog of Galactic Carbon Stars (CGCS; \citealt{Alksnis01}).}
\tablenotetext{d}{Probability of belonging to the RCB class when using the augmented training set.}
\tablenotetext{e}{Relative rank of $P({\rm RCB})$ when using the augmented training set.}
\label{tab:new_cands}
\end{deluxetable*}%

As mentioned in \S~\ref{sec:ML}, one of the major strengths of ML classification is that new discoveries may be fed back 
into the machinery in order to improve future iterations of the classifier. In an attempt to recover more RCB stars and 
DYPers in the ACVS that were missed in our initial search of the MACC, we created an augmented RCB training set by adding 
the eight new RCB stars and DYPers identified in this paper to the 17 sources already included in the training set. This 
augmented training set should increase the likelihood of discovering new candidates, particularly DYPers, of which there 
were no examples in the original training set. 

\begin{figure*}
\begin{center}
\includegraphics[width=130mm]{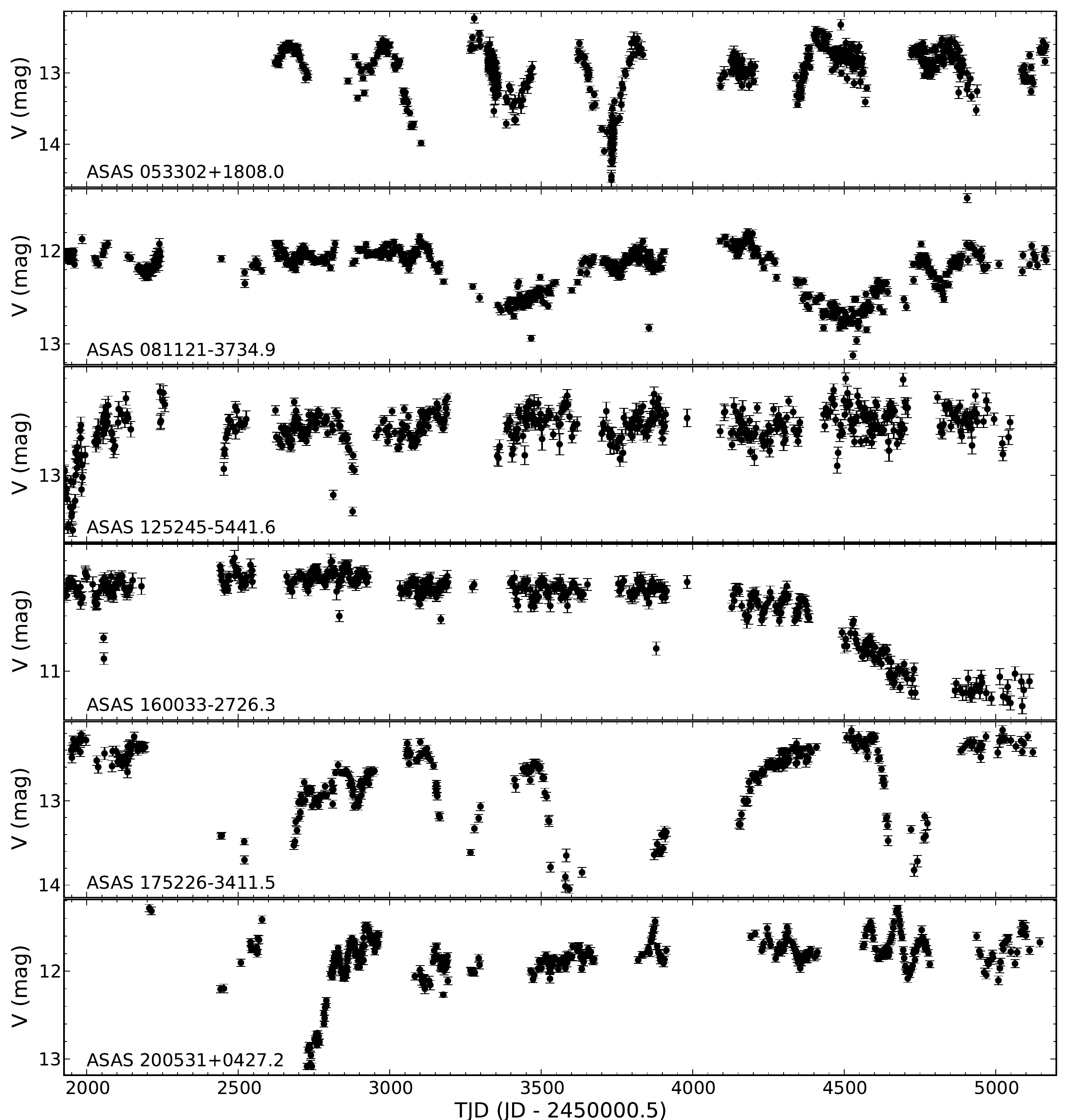}
\caption{ASAS $V$-band light curves of new RCB candidates found using an augmented training set as described in \S~\ref{sec:aug_train}.}
\label{fig:aug_lc}
\end{center}
\end{figure*}

Using the augmented training set we re-ran 
the RF classifier from \citet{Richards12b} on all the ACVS light curves to search for any additional good candidates. We focus 
our new search on candidates with a significant change in $R({\rm RCB})$, which were not 
examined in the initial search of the MACC. In particular, we visually examine the light curves of all sources with 
$R({\rm RCB})_{\rm new} < 500$, $P({\rm RCB})_{\rm MACC} < 0.1$, and $R({\rm RCB})_{\rm new} < R({\rm RCB})_{\rm MACC}$. 
There are a total of 96 sources that meet these criteria, which were not included in the 472 visually inspected sources 
from the original MACC. Of these 96, we conservatively select six as candidate RCB stars or DYPers. One is a highly likely RCB 
star with multiple declines and asymmetric recoveries, three show evidence for a single decline which is only partially 
sampled, and two are known carbon stars that are likely semi-regular periodic variables. We list the candidates in 
Table~\ref{tab:new_cands} with brief comments on each and show their light curves in Figure~\ref{fig:aug_lc}.

\subsection{Future Improvements to the Classifier}

Restricting our search for bright RCB stars to only those sources in the ACVS has biased the results of our search. As was 
mentioned in \S~\ref{sec:training_set}, there are seven known RCB stars that show clear variability in their ASAS light curves 
yet were not selected for inclusion in the ACVS. This suggests that several large-amplitude ASAS variables are missing from 
the ACVS, presumably including a few unknown RCB stars. This bias can easily be corrected by searching all of ASAS for RCB 
stars, however, such a search would include significant new challenges as there are $\sim$12 million sources in ASAS. In 
addition, our existing classification framework is not designed to deal with a catalog where the overwhelming 
majority of sources are not in fact variable. Nevertheless, both of these challenges must be addressed prior to the LSST 
era. We have developed frameworks that can ingest millions of light curves and are currently experimenting with methods 
to deal with non-variable sources, the results of which will be presented in a catalog with classifications for all 
$\sim$12 million sources in ASAS (Richards et al., in prep). Furthermore, 
it has been shown that the use of mid-infrared colors is a powerful discriminant when trying to select RCB stars 
\citep{Tisserand11,Tisserand12}. While our use of NIR colors is important for selecting RCB stars (see e.g., 
\citealt{Soszynski09}), adding the mid-infrared measurements from the all-sky {\it WISE} survey will dramatically improve 
our purity when selecting RCB candidates as several of the Mira and semi-regular variables that served as interlopers in 
the current search (\S~\ref{sec:macc_cands}) would be eliminated with the use of mid-infrared colors.

\section{Conclusions}

We have used the 71-feature Random Forest machine-learning ACVS classification catalog from \citet{Richards12b} to 
identify likely DYPers and RCB stars in the ACVS catalog. The RF classifier provides several advantages over  
previous methods to search time-domain survey data for RCB stars and DYPers. Previously successful searches for RCB stars have 
developed a methodology focused on large 
amplitude variables that do not show strong evidence for periodicity (e.g., \citealt{Alcock01,Zaniewski05,Tisserand08}). While 
the RF classifier is capable of capturing the large variations and irregular declines observed in RCB stars, the use of many 
features allows complex behavior, such as the shape of the decline and recovery, to be captured as well. Another advantage 
of RF classification is that it does not 
require hard cuts on any individual light curve feature, which can exclude real RCB stars from the final candidate selection. 
There are a total of 472 stars with $P({\rm RCB}) > $ 0.1 in version 2.3 of the MACC, 15 of which were selected as 
good RCB or DYPer candidates after visual inspection and existing spectroscopic information. 

Following spectroscopic observations eight of the good candidates were identified as bona fide RCB stars or DYPers. Four of these, 
ASAS 170541$-$2650.1 (GV Oph), ASAS 162229$-$4835.7 (IO Nor), ASAS 165444$-$4925.9, and ASAS 203005$-$6208.0 were confirmed 
as new RCB stars on the basis of (i) their light curves showing irregular, sharp declines of large amplitude ($\Delta{m}_V$ 
$\gg$ 1 mag), (ii) carbon rich spectra showing a lack of evidence for H and $^{13}$C, and (iii) the mid-infrared excess 
observed in their SEDs. Four of the candidates, ASAS 191909$-$1554.4 (V1942 Sgr), ASAS 065113+0222.1, ASAS 162232$-$5349.2, 
and ASAS 182658+0109.0 appear to be Galactic DYPers on the basis of (i) shallow, symmetric declines in their light curves 
occurring at irregular intervals, (ii) carbon rich spectra resembling carbon N stars showing $^{13}$C and weak or no H, 
and (iii) SEDs that show a single peak, but which are too broad to be explained via a single temperature blackbody. With the 
exception of ASAS 170541$-$2650.1, all of the new candidates show evidence for periodic variability near maximum light. We 
incorporate the newly confirmed RCB stars and DYPers into the training set to identify six new candidates as likely RCB stars. 

Our effort has increased the number of known Galactic DYPers from two to six. While the sample size is small, it appears 
that DYPers have pulsations with period $P > 100$ days at maximum light, which is longer than the typical timescale for 
pulsation in RCB stars (see also \citealt{Alcock01}). Each of the new RCB stars 
and DYPers is bright, $V_{\rm max} \lesssim$ 12 mag, which will enable high-resolution spectroscopy for future studies of 
the detailed abundances of these stars. This is particularly important in the case of the four new DYPers, as DY 
Per itself is the only member of the class which has been observed at high resolution to confirm the lack of H absorption 
in the spectrum \citep{Keenan97,Zacs07}. If these stars are shown to be H deficient, it would be strong evidence that DYPers 
are the cool ($T_{\rm eff} \sim$3500 K) analogs to RCB stars. 

We view the results presented herein as one culmination of a broader effort to extract novel science from the time-domain 
survey data deluge. Earlier work focused on determining the most suitable ML frameworks for classification and 
subsequent classification efficiency (see \citealt{Bloom11} for review). While production of ML-based catalogs (e.g., ACVS; 
\citealt{Pojmanski00}) have been the norm for over a decade, we know of no concerted effort to validate the predictions of 
those catalogs. Now having a probabilistic catalog of variable sources \citep{Richards12b} to work with, we can select our 
demographic priors on classes of interest and also decide just how many false-positives we are willing to tolerate in the 
name of improved efficiency. In the case of the construction of a new set of very common stars (e.g., RR Lyrae catalog), 
we might be willing to tolerate a reduced discovery efficiency to preserve a high level of purity. Management of the 
available resources to follow-up the statements made in a probabilistic catalog becomes the next challenge. We were 
obviously most interested in finding new exemplars of two rare classes and thus tolerated a high impurity. In the discovery 
and characterization of several bright RCB stars and DYPers, the payoff of the efficient use of follow-up resources 
enabled by probabilistic classification is evident.

The classification taxonomy of variable stars clearly conflates phenomenology (e.g., ``periodic'') within a physical 
understanding (``pulsating'') of the origin of what is observed.  And while phenomenologically based mining around an envelope 
of class prototypes can turn up new class members, we have shown that the diversity of RCB stars and DYPers demands an 
expanded approach to discovery.  We speculate that the richness and connections of the feature set in the ML search may be 
also capturing some of the phenomenological manifestations of the underlying physics, however nuanced, that we cannot (yet) 
express.

\acknowledgments 

We thank the anonymous referee for comments that have helped to improve this paper.
We thank Alex Filippenko for obtaining the Lick telescope time, while he and Peter Nugent 
contributed to the Keck proposal and assisted in the observations. We thank K.~Clubb, 
A.~Morgan, D.~Cohen and I.~Shivvers for assisting in the Keck and Lick observations. 
We thank Fred Walter for help with the scheduling and reduction of RC spectrograph 
data, and we thank R.\ Hernandez, A.\ Miranda, and M.\ Hernandez for carrying out 
the RC spectrograph observations. 

A.A.M. is supported by the National Science Foundation (NSF) Graduate
Research Fellowship Program. 
J.S.B. and J.W.R. acknowledge support of an NSF-CDI grant-0941742. Some of the work 
for this study was performed in the CDI-sponsored Center for Time Domain Informatics. 
A.A.M. was partially supported by NSF-AAG grant-1009991. 
S.B.C.~acknowledges generous financial assistance from Gary \& Cynthia Bengier, the 
Richard \& Rhoda Goldman Fund, NASA/{\it Swift} grants NNX10AI21G and GO-7100028, 
the TABASGO Foundation, and NSF grant AST-0908886.
Support for K.G.S. is through the Vanderbilt Initiative in Data-intensive Astrophysics (VIDA).

Some of the data presented herein were obtained at the W. M. Keck
Observatory, which is operated as a scientific partnership among the
California Institute of Technology, the University of California, and
NASA. The Observatory was made possible by the generous financial
support of the W. M. Keck Foundation. The authors wish to recognize
and acknowledge the very significant cultural role and reverence that
the summit of Mauna Kea has always had within the indigenous Hawaiian
community.  We are most fortunate to have the opportunity to conduct
observations from this mountain.

This research has made use of NASA's Astrophysics Data System
Bibliographic Services, the SIMBAD database operated at CDS,
Strasbourg, France, the NASA/IPAC Extragalactic Database and 
NASA/ IPAC Infrared Science Archive operated by
the Jet Propulsion Laboratory, California Institute of Technology,
under contract with NASA, and the VizieR database of astronomical
catalogs \citep{Ochsenbein00}. 
Feature computations and classifier research and evaluations were
performed using IBM-\href{http://citris-uc.org/}{CITRIS}'s 280 core
Linux cluster located at UC Berkeley. 

The Digitized Sky Surveys were produced at the Space Telescope Science Institute under U.S. 
Government grant NAG W-2166. The images of these surveys are based on photographic data 
obtained using the Oschin Schmidt Telescope on Palomar Mountain and the UK Schmidt 
Telescope. The plates were processed into the present compressed digital form with 
the permission of these institutions.

{\it Facilities:} 
\facility{ASAS}, \facility{CTIO:1.5m (RC spectrograph)}, \facility{Shane (Kast Double spectrograph)}, \facility{Keck:I (LRIS)}

\end{document}